\documentclass[runningheads,fleqn]{svmult}

\usepackage{makeidx}   
\usepackage{graphicx}  
\usepackage{subeqnar}  
\usepackage{multicol}  
\usepackage{physmult}  
\makeindex             

\usepackage{amsmath} 
\usepackage{amssymb}
\def\lsim{\hbox{ \raise.35ex\rlap{$<$}\lower.6ex\hbox{$\sim$}\ }}
\def\gsim{\hbox{ \raise.35ex\rlap{$>$}\lower.6ex\hbox{$\sim$}\ }}
\def\etal{{\it et al.}}
 \def\eg{{\it e.g.}}  \def\ie{{\it i.e.}}

\def\xrightarrow#1#2#3#4{\,\lower#1pt\hbox{$\stackrel{\stackrel{\displaystyle #2}%
{\hbox to #3cm{\rightarrowfill}}}{#4}$}\,}

\begin{document}
\title*{Cosmic Strings}
%
%
%
%
%
\author{Mairi Sakellariadou} \institute{Department of Physics, King's
College London, University of London, Strand WC2R 2LS, United Kingdom   }
\authorrunning{Mairi Sakellariadou}
\maketitle 
\vskip5mm 

Cosmic strings, a hot subject in the 1980's and early 1990's, lost its
appeal when it was found that it leads to inconsistencies in the power
spectrum of the measured cosmic microwave background temperature
anisotropies. However, topological defects in general, and cosmic
strings in particular, are deeply rooted in the framework of grand
unified theories. Indeed, it was shown that cosmic strings are
expected to be generically formed within supersymmetric grand unified
theories. This theoretical support gave a new boost to the field of
cosmic strings, a boost which has been recently enhanced when it was
shown that cosmic superstrings (fundamental or one-dimensional
Dirichlet branes) can play the r\^ole of cosmic strings, in the
framework of braneworld cosmologies.

To build a cosmological scenario we employ high energy physics;
inflation and cosmic strings then naturally appear. Confronting the
predictions of the cosmological scenario against current
astrophysical/cosmological data we impose constraints on its free
parameters, obtaining information about the high energy physics we
employed.

This is a beautiful example of the rich and fruitful interplay between
cosmology and high energy physics.

\section{Introduction}

The basic ingredient in cosmology is general relativity and the choice
of a metric. The Friedmann-Lema\^{i}tre-Robertson-Walker (FLRW) model,
known as the {\sl hot big bang} model is a homogeneous and isotropic
solution of Einstein's equations; the hyper-surfaces of constant time
are homogeneous and isotropic, \ie, spaces of constant curvature.  The
hot big bang model is based on the FLRW metric
\begin{equation}
ds^2=dt^2-a^2(t)\gamma_{ij}dx^i dx^j=
a^2(\tau)[d\tau^2-\gamma_{ij}dx^idx^j]~;
\end{equation} 
$a(t)$ or $a(\tau)$ is the cosmic scale factor in terms of the
cosmological time $t$ or the conformal time $\tau$ (with $dt=ad\tau$)
respectively, and $\gamma_{ij}$ is the metric of a space with constant
curvature $\kappa$. The metric $\gamma_{ij}$ can be expressed as
\begin{eqnarray}
 \gamma_{ij} dx^i dx^j &=&
dr^2+\chi^2(r)\left(d\vartheta^2+sin^2\vartheta d\varphi^2\right)~,\\
\mbox{with}~ ~ \chi(r) &=& \left\{ \begin{array}{lcl} r & \ \
\mbox{for}\ \ & \kappa=0 \\ \sin r & \ \ \mbox{for}\ \ & \kappa=1 \\
\sinh r & \ \ \mbox{for}\ \ & \kappa=-1 ~; \end{array} \right.
\end{eqnarray}
the scale factor $a(\tau)$ has been rescaled so that the curvature is
$\kappa=\pm 1$ or $0$.

The cornerstone of the hot big bang model is the high degree of
symmetry of the FLRW metric: there is only one dynamical variable, the
cosmic scale factor. The FLRW model is so successful that became the
standard cosmological model. The high degree of symmetry of the
metric, originally a theorist's simplification, is now an evidence
thanks to the remarkable uniformity of temperature of the Cosmic
Microwave Background (CMB) measured first by the COBE-DMR
satellite~\cite{cobe}.

The four pillars upon which the success of standard hot big bang model
lie are: (i) the expansion of the Universe, (ii) the origin of the
cosmic background radiation, (iii) the synthesis of light elements,
and (iv) the formation of large-scale structures. However, there are
questions, which mainly concern the initial conditions, to which the
hot big bang model is unable to provide an answer. These shortcomings
of the FLRW model are: (i) the horizon problem, (ii) the flatness
problem, (iii) the exotic relics, (iv) the origin of density
fluctuations, (v) the cosmological constant, and (vi) the singularity
problem.  To address these issues, inflation was
proposed~\cite{infl1,infl2}.  Inflation essentially consists of a
phase of accelerated expansion, corresponding to repulsive gravity and
an equation of state $3p<-\rho$, which took place at a very high
energy scale. Even though inflation is at present the most appealing
scenario to describe the early stages of the Universe, the issue of
how generic is the onset of inflation is still under
discussion~\cite{ecms,gt,us}, at least within a large class of
inflationary potentials, in the context of classical general relativity
and loop quantum cosmology.

From the observational point of view, the remarkable uniformity of the
CMB indicates that at the epoch of last scattering, approximately
$2\times 10^5 {\rm yr}$ after the big bang, when the Universe was at a
temperature of approximately $0.26 {\rm eV}\simeq 3\times 10^3 {\rm
K}$, the Universe was to a high degree or precision ($10^{-5}$)
isotropic and homogeneous.  At very large scales, much bigger than
$110 {\rm Mpc}\approx 10^{21} {\rm km}$, the Universe is smooth, while
at small scales the Universe is very lumpy. The fractional overdensity
at the time of decoupling between baryons and photons was
\begin{equation}
\biggl(\frac{\delta\rho}{\rho}\biggr)_{\rm dec}={\cal C}\times
\biggl(\frac{\delta T}{T}\biggr)\leq {\cal O}(10^{-2}-10^{-3})~;
\end{equation}
the constant ${\cal C}$ depends on the nature of density perturbations
and it is ${\cal C}={\cal O}(10-100)$. Then one asks the following
question: {\sl how does a very smooth Universe at the time of
decoupling became very lumpy today?}

In the 1980's and 1990's, cosmologists had the following picture in
mind: small, primeval density inhomogeneities grew via gravitational
instability into the large inhomogeneities we observe today. To
address the question of the origin of the initial density
inhomogeneities, one needs to add more ingredients (namely scalar
fields) to the cosmological model. This is where high energy physics
enter the picture. Clearly, to build a detailed scenario of structure
formation one should know the initial conditions, \ie, the total
amount of nonrelativistic matter, the composition of the Universe, the
spectrum and type of primeval density perturbations.

For almost two decades, two families of models have been considered
challengers for describing, within the framework of gravitational
instability, the formation of large-scale structure in the Universe.
Initial density perturbations can either be due to {\sl freezing in}
of quantum fluctuations of a scalar field during an inflationary
period, or they may be seeded by a class of topological
defects~\cite{td1}, which could have formed naturally during a
symmetry breaking phase transition in the early Universe.  On the one
hand, quantum fluctuations amplified during inflation produce {\sl
adiabatic}, or {\sl curvature} fluctuations with a scale-invariant
spectrum. It means that there are fluctuations in the local value of
the spatial curvature, and that the fractional overdensity in Fourier
space behaves as $|\delta_k|^2\propto k^{-3}$. If the quantum
fluctuations of the inflaton field are in the vacuum state, then the
statistics of the CMB is Gaussian~\cite{jam,gms}.  On the other hand,
topological defects trigger {\sl isocurvature}, or {\sl isothermal}
fluctuations, meaning that there are fluctuations in the form of the
local equation of state, with nongaussian statistics and a
scale-invariant spectrum.  The CMB anisotropies provide a link between
theoretical predictions and observational data, which may allow us to
distinguish between inflationary models and topological defects
scenarios, by purely linear analysis.  The characteristics of the CMB
anisotropy multipole moments (position, amplitude of acoustic peaks),
and the statistical properties of the CMB are used to discriminate
among models, and to constrain the parameters space.

Many particle physics models of matter admit solutions which
correspond to a class of topological defects, that are either stable
or long-lived.  Provided our understanding about unification of forces
and the big bang cosmology are correct, it is natural to expect that
such topological defects could have formed naturally during phase
transitions followed by spontaneously broken symmetries, in the early
stages of the evolution of the Universe.  Certain types of topological
defects (local monopoles and local domain walls) lead to disastrous
consequences for cosmology and thus, they are undesired, while others
may play a useful r\^ole.  We consider gauge theories, thus we are
only interested in comic strings, since on the one hand strings are
not cosmologically dangerous (monopoles and domain walls are), and on
the other hand they can be useful in cosmology (textures decay too
fast).

Cosmic strings are linear topological defects, analogous to flux tubes
in type-II superconductors, or to vortex filaments in superfluid
helium.  In the framework of Grand Unifies Theories (GUTs), cosmic
strings might have been formed at a grand unification transition, or
much later, at the electroweak transition, or even at an intermediate
one. These objects carry a lot of energy and they could play a r\^ole
in cosmology and/or astrophysics. In the simplest case, the linear
mass density of a cosmic string, denoted by $\mu$, is equal to the
string tension. Thus, the characteristic speed of waves on the string
is the speed of light. For strings produced at a phase transition
characterised by temperature $T_{\rm c}$, one expects roughly $\mu\sim
T_{\rm c}^2$. The strength of the gravitational interaction of cosmic
strings is given in terms of the dimensionless quantity $G\mu\sim
(T_{\rm c}/M_{\rm Pl})^2$, where $G$ and $M_{\rm Pl}$ denote the
Newton's constant and the Planck mass, respectively. For grand
unification strings, the energy per unit length is $\mu\sim 10^{22}
{\rm kg}/{\rm m}$, or equivalently, $G\mu\sim {\cal
O}(10^{-6}-10^{-7})$.

Topological defects (global or local) in general, and cosmic strings
in particular, are ruled out as the unique source of the measured CMB
temperature anisotropies. Clearly, one should then address the
following question: {\sl which are the implications for the high
energy physics models upon which the cosmological scenario is based?}
This leads to the following list of questions: (i) how generic is
cosmic string formation? (ii) which is the r\^ole of cosmic strings,
if any? and (iii) which is a natural inflationary scenario (inflation
is still a paradigm in search of a model)? These questions will be
addressed in what follows. We will see that cosmic strings are
generically formed at the end of an inflationary era, within the
framework of Supersymmetric Grand Unified Theories (SUSY GUTs).  This
implies that cosmic strings have to be included as a sub-dominant
partner of inflation. We will thus consider mixed models, where both
the inflaton field and cosmic strings contribute to the measured CMB
temperature anisotropies. Comparing theoretical predictions against
CMB data we will find the maximum allowed contribution of cosmic
strings to the CMB measurements. We will then ask whether the free
parameters of supersymmetric inflationary models can be adjusted so
that the contribution of strings to the CMB is within the allowed
window.

Finally, the recent proposal that cosmic superstrings can be
considered as cosmic string candidates opens new perspectives on the
theoretical point of view. More precisely, in the framework of large
extra dimensions, long superstrings may be stable and appear at the
same energy scale as GUT scale cosmic strings.

In what follows we first discuss, in Section {\bf 2}, topological
defects in GUTs. We classify topological defects and we give the
criterion for their formation. We then briefly discuss two simple
models leading to the formation of global strings (vortices) and local
(gauge) strings, namely the Goldstone and the abelian-Higgs model,
respectively. Next, we present the Kibble and Zurek mechanisms of
topological defect formation. We concentrate on local gauge strings
(cosmic strings) and give the equations of motion for strings in the
limit of zero thickness, moving in a curved spacetime. We subsequently
discuss the evolution of a cosmic string network; the results are
based on heavy numerical simulations. We then briefly present string
statistical mechanics and the Hagedorn phase transition. We end this
Section by addressing the question of whether cosmic strings are
expected to be generically formed after an inflationary era, in the
context of supersymmetric GUTs.  In Section {\bf 3} we discuss the
most powerful tool to test cosmological predictions of theoretical
models, namely the spectrum of CMB temperature anisotropies. We then
analyse the predictions of models where the initial fluctuations
leading to structure formation and the induced CMB anisotropies were
triggered by topological defects.  In Section {\bf 4}, we study
inflationary models in the framework of supersymmetry and
supergravity.  In Section {\bf 5} we address the issue of cosmic
superstrings as cosmic strings candidates, in the context of
braneworld cosmologies. We round up with our conclusions in Section
{\bf 6}.

\section{Topological Defects}

\subsection{Topological Defects in GUTs}

The Universe has steadily cooled down since the Planck time, leading
to a series of Spontaneously Symmetry Breaking (SSB), which may lead
to the creation of topological defects~\cite{td1}, false vacuum
remnants, such as domain walls, cosmic strings, monopoles, or
textures, via the Kibble mechanism~\cite{kibble}.

The formation or not of topological defects during phase transitions,
followed by SSB, and the determination of the type of the defects,
depend on the topology of the vacuum manifold ${\cal M}_n$.  The
properties of ${\cal M}_n$ are usually described by the $k^{\rm th}$
homotopy group $\pi_k({\cal M}_n)$, which classifies distinct mappings
from the $k$-dimensional sphere $S^k$ into the manifold ${\cal
M}_n$. To illustrate that, let us consider the symmetry breaking of a
group G down to a subgroup H of G . If ${\cal M}_n={\rm G}/{\rm H}$
has disconnected components, or equivalently if the order $k$ of the
nontrivial homotopy group is $k=0$, then two-dimensional defects,
called {\sl domain walls}, form.  The spacetime dimension $d$ of the
defects is given in terms of the order of the nontrivial homotopy
group by $d=4-1-k$. If ${\cal M}_n$ is not simply connected, in other
words if ${\cal M}_n$ contains loops which cannot be continuously
shrunk into a point, then {\sl cosmic strings} form. A necessary, but
not sufficient, condition for the existence of stable strings is that
the first homotopy group (the fundamental group) $\pi_1({\cal M}_n)$
of ${\cal M}_n$, is nontrivial, or multiply connected. Cosmic strings
are line-like defects, $d=2$. If ${\cal M}_n$ contains unshrinkable
surfaces, then {\sl monopoles} form, for which $k=1, ~d=1$.  If ${\cal
M}_n$ contains noncontractible three-spheres, then event-like defects,
{\sl textures}, form for which $k=3, ~d=0$.

Depending on whether the symmetry is local (gauged) or global (rigid),
topological defects are called {\sl local} or {\sl global}. The energy
of local defects is strongly confined, while the gradient energy of
global defects is spread out over the causal horizon at defect
formation.  Patterns of symmetry breaking which lead to the formation
of local monopoles or local domain walls are ruled out, since they
should soon dominate the energy density of the Universe and close it,
unless an inflationary era took place after their formation.  Local
textures are insignificant in cosmology since their relative
contribution to the energy density of the Universe decreases rapidly
with time~\cite{textures}.

Even if the nontrivial topology required for the existence
of a defect is absent in a field theory, it may still be possible to have
defect-like solutions. Defects may be {\sl embedded}
in such topologically trivial field theories~\cite{embedded}. While
stability of topological defects is guaranteed by topology, embedded
defects are in general unstable under small perturbations.

\subsection{Spontaneous Symmetry Breaking}

The concept of Spontaneous Symmetry Breaking has its origin in
condensed matter physics. In field theory, the r\^ole of the order
parameter is played by scalar fields, the Higgs fields. The symmetry is
said to be spontaneously broken if the ground state is characterised
by a nonzero expectation value of the Higgs field and does not exhibit
the full symmetry of the Hamiltonian.

\subsubsection{The Goldstone Model} 

To illustrate the idea of SSB we 
consider the simple Goldstone model. Let $\phi$ be a complex scalar
field with classical Lagrangian density
\begin{equation}
{\cal L}=(\partial_\mu\bar\phi)(\partial^\mu\phi)-V(\phi)~,
\label{lgm}
\end{equation}
and potential $V(\phi)$:
\begin{equation}
V(\phi)=\frac{1}{4}\lambda[\bar\phi\phi-\eta^2]^2~,
\label{mh}
\end{equation}
with positive constants $\lambda,\eta$. This potential, Eq.~(\ref{mh}),
has the symmetry breaking {\itshape Mexican hat} shape.
The Goldstone model is invariant under the U(1) group of global phase
transformations,
\begin{equation}
\phi(x)\rightarrow e^{i\alpha}\phi(x)~,
\label{ptg}
\end{equation}
where $\alpha$ is a constant, \ie,  independent of spacetime.  The
minima of the potential, Eq.~(\ref{mh}), lie on a circle with fixed
radius $|\phi|=\eta$; the ground state of the theory is characterised
by a nonzero expectation value, given by
\begin{equation}
\langle 0|\phi|0\rangle =\eta e^{i\theta}~,
\end{equation}
where $\theta$ is an arbitrary phase.  The phase transformation,
Eq.~(\ref{ptg}), leads to the change $\theta\rightarrow
\theta+\alpha$, which implies that the vacuum state $|0\rangle$ is not
invariant under the phase transformation, Eq.~(\ref{ptg}); the
symmetry is spontaneously broken.  The state of unbroken symmetry with
$\langle 0|\phi|0\rangle =0$ is a local maximum of the Mexican hat
potential, Eq.~(\ref{mh}). All broken symmetry vacua, each with a
different value of the phase $\theta$ are equivalent. Therefore, if we
select the vacuum with $\theta=0$, the complex scalar field $\phi$ can
be written in terms of two real scalar fields, $\phi_1,\phi_2$, with
zero vacuum expectation values, as
\begin{equation}
\phi=\eta+\frac{1}{\sqrt 2}(\phi_1+i\phi_2)~.
\end{equation}
As a consequence, the Lagrangian density, Eq.~(\ref{lgm}), can be
written as
\begin{equation}
{\cal L}=\frac{1}{2}(\partial_\mu \phi_1)^2+\frac{1}{2}(\partial_\mu
\phi_2)^2-\frac{1}{2}\lambda\eta^2\phi_1^2+{\cal L}_{\rm int}~.
\end{equation}
The last term, ${\cal L}_{\rm int}$, is an interaction term which
includes cubic and higher-order terms in the real scalar fields
$\phi_1,\phi_2$. Clearly, $\phi_1$ corresponds to a massive particle,
with mass $\sqrt{\lambda}\eta >0$, while $\phi_2$ corresponds to a
massless scalar particle, the Goldstone boson. The appearance of
Goldstone bosons is a generic feature of models with spontaneously
broken global symmetries.

Going around a closed path $L$ in physical space, the phase $\theta$
of the Higgs field $\phi$ develops a nontrivial winding, \ie,
$\Delta\theta=2\pi$. This closed path can be shrunk continuously to a
point, only if the field $\phi$ is lifted to the top of its potential
where it takes the value $\phi=0$. Within a closed path for which the
total change of the phase of the Higgs field $\phi$ is $2\pi$, a
string is trapped.  A string must be either a closed loop or an
infinitely long (no ends) string, since otherwise one could deform the
closed path $L$ and avoid to cross a string.

We should note that we considered above a purely classical potential,
Eq.~(\ref{mh}), to determine the expectation value of the Higgs field
$\phi$. In a more realistic case however, the Higgs field $\phi$ is a
quantum field which interacts with itself, as well as with other
quantum fields. As a result the classical potential $V(\phi)$ should
be modified by radiative corrections, leading to an effective
potential $V_{\rm eff}(\phi)$. There are models for which the
radiative corrections can be neglected, while there are others for
which they play an important r\^ole.

The Goldtone model is an example of a second-order phase transition
leading to the formation of global strings, {\sl vortices}.

\subsubsection{The Abelian-Higgs Model} 

We are interested in local (gauge) strings (cosmic strings), so let us
consider the simplest gauge theory with a spontaneously broken
symmetry. This is the abelian-Higgs model with Lagrangian density
\begin{equation}
{\cal L}=\bar{{\cal} D}_\mu\phi{\cal D}^\mu\phi-\frac{1}{4}F_{\mu\nu}
F^{\mu\nu}-V(\phi)~, 
\label{lahm}
\end{equation}
where $\phi$ is a complex scalar field with potential $V(\phi)$, given
by Eq.~(\ref{mh}), and $F_{\mu\nu}=\partial_\mu A_\nu-\partial_\nu
A_\mu$ is the field strength tensor. The covariant derivative ${\cal
D}_\mu$ is defined by ${\cal D}_\mu=\partial_\mu-ieA_\mu$, with $e$
the gauge coupling constant and $A_\mu$ the gauge field.  The
abelian-Higgs model is invariant under the group U(1) of local gauge
transformations
\begin{equation}
\phi(x)\rightarrow  e^{i\alpha(x)}\phi(x)~ ~ ; ~ ~ A_\mu(x)\rightarrow
A_\mu(x)+\frac{1}{e}\partial_\mu \alpha(x)~,
\end{equation}
where $\alpha(x)$ is a real single-valued function. 

The minima of the Mexican hat potential, Eq.~(\ref{mh}), lie on a
circle of fixed radius $|\phi|=\eta$, implying that the symmetry is
spontaneously broken and the complex scalar field $\phi$ acquires a
nonzero vacuum expectation value. Following the same approach as in
the Goldstone model, we chose to represent $\phi$ as
\begin{equation} 
\phi=\eta+\frac{\phi_1}{\sqrt{2}}~,
\end{equation}
leading to the Lagrangian density
\begin{equation}
{\cal L}=\frac{1}{2}(\partial_\mu\phi_1)^2-\frac{1}{2}\mu^2\phi_1^2
-\frac{1}{4}F_{\mu\nu}F^{\mu\nu}+\frac{1}{2}M^2 A_\mu A^\mu+{\cal
  L}_{\rm int}~,
\end{equation}
where the particle spectrum contains a scalar particle (Higgs boson)
with mass $m_{\rm s}=\sqrt{\lambda}\eta$ and a vector field (gauge
boson) with mass $m_{\rm v}=\sqrt{2}e\eta$. The breaking of a
gauge symmetry does not imply a massless Goldstone boson.  The
abelian-Higgs model is the simplest model which admits string
solutions, the Nielsen-Olesen vortex lines.  The width of the string
is determined by the Compton wavelength of the Higgs and gauge bosons,
which is $\sim m_{\rm s}^{-1}$ and $\sim m_{\rm v}^{-1}$,
respectively.

In the Lorentz gauge, $\partial_\mu A^\mu=0$, the Higgs field $\phi$
has the same form as in the case of a global string at large distances
from the string core, \ie,
\begin{equation}
\phi\approx\eta e^{in\theta}~,
\label{aff}
\end{equation}
where the integer $n$ denotes the string winding number.  The gauge
field asymptotically approaches
\begin{equation}
A_\mu\approx\frac{1}{ie}\partial_\mu\ln\phi~.
\label{afa}
\end{equation}
The asymptotic forms for the Higgs and gauge fields, Eqs.~(\ref{aff})
and (\ref{afa}) respectively, imply that far from the string core, we
have
\begin{equation}
{\cal D}_\mu\phi\approx 0 ~ ~ , ~ ~ F_{\mu\nu}\approx 0~.
\end{equation}
As a consequence, far from the string core, the energy density
vanishes exponentially, while the total energy per unit length is
finite. The string linear mass density $\mu$ is
\begin{equation}
\mu\sim\eta^2~.
\end{equation}
In the case of a global U(1) string there is no gauge field to
compensate the variation of the phase at large distances from the
string core, resulting to a linear mass density which diverges at long
distances from the string. For a global U(1) string with winding
number $n=1$ one obtains
\begin{equation}
\mu\sim\eta^2+\int_\delta^R\biggl[\frac{1}{r}\frac{\partial\phi}
{\partial\theta}\biggr]^2 2\pi
rdr\approx2\pi\eta^2\ln\biggl(\frac{R}{\delta}\biggr)~,
\end{equation} 
where $\delta$ stands for the width of the string core and $R$ is a
cut-off radius at some large distance from the string, \eg~ the
curvature radius of the string, or the distance to the nearest string
segment in the case of a string network. The logarithmic term in the
expression for the string energy mass density per unit length leads to
long-range interactions between global U(1) string segments, with a
force $\sim\eta^2/R$.

The field equations arising from the Lagrangian density,
Eq.~(\ref{lahm}), read
\begin{eqnarray}
(\partial_\mu-ie A_\mu)(\partial^\mu-ie
A^\mu)\phi+\frac{\lambda}{2}\phi (\phi\bar\phi-\eta^2)&=&0\nonumber\\
\partial_\nu F^{\mu\nu}-2e {\rm
Im}\left[{\bar\phi}(\partial^\nu-ieA^\nu)\phi \right]&=&0~.
\end{eqnarray} 
The equations of motion can be easily solved for the case of straight,
static strings.

The internal structure of the string is meaningless when we deal with
scales much larger than the string width. Thus, for a straight string
lying along the $z$-axis, the effective energy-momentum tensor is
\begin{equation}
{\tilde T}^\mu_{\ \nu}=\mu\delta(x)\delta(y){\rm diag}(1,0,0,1)~.
\label{eemt}
\end{equation}

\subsection{Thermal Phase Transitions and Defect Formation}

In analogy to condensed matter systems, a symmetry which is
spontaneoulsy broken at low temperatures can be restored at higher
temperatures. In field theories, the expectation value of the Higgs
field $\phi$ can be considered as a Bose condensate of Higgs
particles. If the temperature $T$ is nonzero, one should consider a
thermal distribution of particles/antiparticles, in addition to the
condensate. The equilibrium value of the Higgs field $\phi$ is
obtained by minimising the free energy $F=E-TS$. Only at high enough
temperatures the free energy is effectively temperature-dependent,
while at low temperatures the free energy is minimised by the ordered
state of the minimum energy. 

Let us consider for example the Goldstone model, for which the
high-temperature effective potential is
\begin{equation}
V_{\rm eff}(\phi,T)=m^2(T)|\phi|^2=\frac{\lambda}{4}|\phi|^4~~
\mbox{where}~~m^2(T)=\frac{\lambda}{12}(T^2-6\eta^2)~.
\end{equation}
The effective mass-squared term $m^2(T)$ for the Higgs field $\phi$ in
the symmetric state $\langle\phi\rangle=0$, vanishes at the critical
temperature $T_{\rm c}=\sqrt{6}\eta$. The effective potential is
calculated using perturbation theory and the leading contribution
comes from one-loop Feynman diagrams. For a scalar theory, the main
effect is a temperature-dependent quadratic contribution to the
potential. Above the critical temperature, $m^2(T)$ is positive,
implying that the effective potential gets minimised at $\phi =0$,
resulting to a symmetry restoration. Below the critical temperature,
$m^2(T)$ is negative, implying that the Higgs field has a nonvanishing
expectation value.

Even if there is symmetry restoration and the mean value
$\langle\phi\rangle$ of the Higgs field vanishes, the actual value of
the field $\phi$ fluctuates around the mean value, meaning that $\phi$
at any given point is nonzero. The thermal fluctuations have, to a
leading approximation, a Gaussian distribution, thus they can be
characterised by a two-point correlation function, which typically
decays exponentially, with a decay rate characterised by the {\sl
correlation length} $\xi$. The consequence of this is that
fluctuations at two points separated by a distance greater than the
correlation length $\xi$ are independent.

Kibble~\cite{kibble} was first to estimate the initial density of
topological defects formed after a phase transition followed by SSB in
the context of cosmology. His criterion was based on the causality
argument and the Ginzburg temperature, $T_{\rm G}$, defined as the
temperature below which thermal fluctuations do not contain enough
energy for regions of the field on the scale of the correlation length
to overcome the potential energy barrier and restore the symmetry,
\begin{equation}
\xi^3(T_{\rm G})\Delta F(T_{\rm G})\sim T_{\rm G}~;
\end{equation}
$\Delta F$ is the difference in free energy density between the false
and true vacua.

According to the Kibble mechanism, the initial defect network is
obtained by the equilibrium correlation length of the Higgs field at
the Ginzburg temperature. Consequently, laboratory tests confirmed
defect formation at the end of a symmetry breaking phase transition, but
they disagree with defect density estimated by Kibble. More
precisely, Zurek~\cite{zurek1,zurek2} argued that the the relaxation
time ${\bar{\tau}}(T)$, which is the time it takes correlations to
establish on the length scale $\xi(T)$, has an important r\^ole in
determining the initial defect density. 

Let us describe the {\sl freeze-out} proposal suggested by Zurek to
estimate the initial defect density.  Above the critical temperature
$T_{\rm c}$, the field starts off in thermal equilibrium with a heat
bath. Near the phase transition, the equilibrium correlation length
diverges
\begin{equation}
\xi(T)=\xi_0\biggl(\frac{T-T_{\rm c}}{T_{\rm c}}\biggl)^{-\nu}~,
\end{equation}
where $\nu$ denotes the critical component. At the same time, the
dynamics of the system becomes slower, and this can be expressed in
terms of the equilibrium relaxation timescale of the field, which also
diverges, but with a different exponent $\mu$:
\begin{equation}
{\bar \tau}(T)={\bar \tau}_0\biggl(\frac{T-T_{\rm c}}{T_{\rm
c}}\biggl)^{-\mu}~.
\end{equation}
The values of the critical components $\mu,\nu$ depend on the theory
under consideration. 
Assuming, for simplicity, that the temperature is decreasing linearly,
\begin{equation}
T(t)=\biggl(1-\frac{t}{{\bar \tau}_{\rm Q}}\biggr)T_{\rm c}~,
\end{equation}
where the {\sl quench timescale} ${\bar \tau}_{\rm Q}$ characterises the cooling rate.

As the temperature decreases towards its critical temperature $T_{\rm
c}$, the correlation length $\xi$ grows as
\begin{equation}
\xi(t)\sim \biggl(\frac{|t|}{{\bar \tau}_{\rm Q}}\biggl)^{-\nu}~,
\end{equation}
but at the same time the dynamics of the system becomes slower,
\begin{equation}
{\bar \tau}(t)\sim \biggl(\frac{|t|}{{\bar\tau}_{\rm Q}}\biggl)^{-\mu}~.
\end{equation}
As the system approaches from above the critical temperature, there
comes a time $|\hat t|$ during the quench when the equilibrium
relaxation timescale equals the time that is left before the
transition at the critical temperature, namely
\begin{equation}
{\bar \tau}(\hat t)=|\hat t|~.
\end{equation}
After this time, the system can no longer adjust fast enough to the
change of the temperature of the thermal bath, and falls out of
equilibrium. At time $\hat t$, the dynamics of the correlation length
freezes. The correlation length cannot grow significantly after this
time, and one can safely state that it freezes to its value at time
$\hat t$. Thus, according to Zurek's proposal the initial defect density
is determined by the freeze-out scale~\cite{zurek1,zurek2}
\begin{equation}
\hat\xi\equiv\xi(\hat t)\sim {\bar\tau}_{\rm Q}^{\nu/(1+\mu)}~.
\label{defform}
\end{equation}
We note that the above discussion is in the framework of second-order
phase transitions.

The above prediction, Eq.~(\ref{defform}), has been tested
experimentally in a variety of systems, as for example, in superfluid
$^4$He~\cite{hel4a,hel4b} and $^3$He~\cite{hel3a,hel3b}, and in liquid
crystals~\cite{lca,lcb,lcc}. Apart the experimental support, the
Kibble-Zurek picture is supported by numerical
simulations~\cite{lz,yz,zbz} and calculations using the methods of
nonequilibrium quantum field theory~\cite{schr}.

\subsection{Cosmic String Dynamics}

The world history of a string can be expressed by a two-dimensional
surface in the four-dimensional spacetime, which is called the string
worldsheet:
\begin{equation}
x^\mu=x^\mu(\zeta^a)~~,~~a=0,1~;
\end{equation}
the worldsheet coordinates $\zeta^0, \zeta^1$ are arbitrary parameters
chosen so that $\zeta^0$ is timelike and $\zeta^1$ spacelike ($\equiv
\sigma$).

 The string equations of motion, in the limit of a zero thickness
string, are derived from the Goto-Nambu effective action which, up to
an overall factor, corresponds to the surface area swept out by the
string in spacetime:
\begin{equation}
S_0[x^\mu]=-\mu\int\sqrt{-\gamma}d^2\zeta~,
\label{nga}
\end{equation}
where $\gamma$ is the determinant of the two-dimensional worldsheet
metric $\gamma_{ab}$,
\begin{equation}
\gamma={\rm det}(\gamma_{ab})=\frac{1}{2}
\epsilon^{ac}\epsilon^{bd}\gamma_{ab}\gamma_{cd}~~ ,
~~\gamma_{ab}=g_{\mu\nu}x^\mu_{,a}x^\nu_{,b}~.
\label{gamma}
\end{equation}
If the string curvature is small but not negligible, one may
consider an expansion in powers of the curvature, leading to the
following form for the string action up to second order
\begin{equation}
S=-\int d^2\zeta\sqrt{-\gamma}(\mu-\beta_1 K^A K^A+\beta_2 R)~,
\end{equation}
where $\beta_1, \beta_2$ are dimensionless numbers. The Ricci
curvature scalar $R$ is a function of the extrinsic curvature tensor
$K^A_{ab}$ (with $A=1,2$), 
\begin{equation}
R=K^{abA}K^A_{ab}-K^AK^A~;
\end{equation}
$K^A=\gamma^{ab}K^A_{ab}$.  Finite corrections and their effects to
 the effective action have been studied by a number of
 authors~\cite{fincor1}-\cite{fincor4}.

By varying the action, Eq.~(\ref{nga}), with respect to
$x^\mu(\zeta^a)$, and using the relation $d\gamma=\gamma
\gamma^{ab}d\gamma_{ab}$, where $\gamma_{ab}$ is given by
Eq.~(\ref{gamma}b), one gets the string equations o motion:
\begin{equation}
x^{\mu\
;a}_{,a}+\Gamma^\mu_{\nu\sigma}\gamma^{ab}x^\nu_{,a}x^\sigma_{,b}=0~,
\label{semnb}
\end{equation}
where $\Gamma^\mu_{\nu\sigma}$ is the four-dimensional Christoffel
symbol,
\begin{equation}
\Gamma^\mu_{\nu\sigma}=\frac{1}{2} g^{\mu\tau}(g_{\tau\nu
,\sigma}+g_{\tau\sigma ,\nu}-g_{\nu\sigma ,\tau})~,
\end{equation}
and the covariant Laplacian is
\begin{equation}
x^{\mu\ ;a}_{,a}=\frac{1}{\sqrt{-\gamma}}\partial_a
(\sqrt{-\gamma}\gamma^{ab}x^\mu_{, b})~.
\end{equation}
One can derive the same string equations of motion by using Polyakov's
form of the action~\cite{pol}
\begin{equation}
S[x^\mu, h_{ab}]=-\frac{\mu}{2}\int \sqrt{-h}h^{ab}\gamma_{ab}d^2\zeta~,
\label{spol}
\end{equation}
where $h_{ab}$ is the internal metric with determinant $h$.

Including a force of friction $F^{\mu\nu}$ due to the scattering of
thermal particles off the string, the equation of motion
reads~\cite{friction}
\begin{equation}
\mu\bigl[x^{\mu\
    ;a}_{,a}+\Gamma^\mu_{\nu\sigma}
\gamma^{ab}x^\nu_{,a}x^\sigma_{,b}\bigr]=
F^\mu(u^\lambda_\perp,T,\sigma)~.
\end{equation}
The force of friction depends on the temperature of the surrounding
matter $T$, the velocity of the fluid transverse to the world sheet
$u^\nu_\perp\equiv u^\nu-x^\nu_{,a}x^{\sigma , \alpha}u_\sigma$, and
the type of interaction between the particles and the string, which we
represent by $\sigma$. Cosmic strings of mass per unit length $\mu$
would have formed at cosmological time 
\begin{equation}
t_0\sim(G\mu)^{-1}t_{\rm Pl}~,
\end{equation}
where $t_{\rm Pl}$ is the Planck time. Immediately after the phase
transition the string dynamics would be dominated by
friction~\cite{friction}, until a time of order
\begin{equation}
t_\star\sim(G\mu)^{-2}t_{\rm Pl}~.
\end{equation}
For cosmic strings formed at the grand unification scale, their mass
per unit length is of order $G\mu\sim 10^{-6}$ and friction is
important only for a very short period of time. However, if strings
have formed at a later phase transition, for example closer to the
electroweak scale, their dynamics would be dominated by friction
through most of the thermal history of the Universe.  The evolution of
cosmic strings taking into account the frictional force due to the
surrounding radiation has been studied in Ref.~\cite{jaume-mairi}.

The string energy-momentum tensor can be obtained by varying the
action, Eq.~(\ref{nga}), with respect to the metric $g_{\mu\nu}$,
\begin{equation}
T^{\mu\nu}\sqrt{-g}=-2\frac{\delta S}{\delta g_{\mu\nu}} =\mu\int
d^2\zeta\sqrt{-\gamma}\gamma^{ab}x^\mu_{,a}x^\nu_{,b}
\delta^{(4)}(x^\sigma-x^\sigma(\zeta^a))~.
\end{equation}
For a straight cosmic string in a flat spacetime lying along the
$z$-axis and choosing $\zeta^0=t$, $\zeta^1=z$, the above expression
reduces to the one for the effective energy-momentum tensor,
Eq.~(\ref{eemt}).

\subsubsection{Cosmic Strings in Curved Spacetime}

The equations of motion for strings are most conveniently written in
comoving coordinates, where the FLRW metric takes the form
\begin{equation}
ds^2=a^2(\tau)[d\tau^2-d{\bf r}^2]~.
\end{equation}
The comoving spatial coordinates of the string, ${\bf
x}(\tau,\sigma)$, are written as a function of conformal time $\tau$
and the length parameter $\sigma$. We have thus chosen the gauge
condition $\zeta^0=\tau$.  For a cosmic string moving in a FLRW
Universe, the equations of motion, Eq.~(\ref{semnb}), can be
simplified by also choosing the gauge in which the unphysical parallel
components of the velocity vanish,
\begin{equation}
\dot {\bf x}\cdot {\bf x}'=0~,
\label{gce}
\end{equation}
where overdots denote derivatives with respect to conformal time
$\tau$ and primes denote spatial derivatives with respect to $\sigma$.

In these coordinates, the Goto-Nambu action yields the following
equations of motion for a string moving in a FLRW matric:
\begin{equation}
\ddot {\bf x}+2\biggl(\frac{\dot a}{a}\biggr)\dot {\bf x} (1-\dot {\bf
x}^2)=\biggl(\frac{1}{\epsilon}\biggr)\biggl(\frac{{\bf
x}'}{\epsilon}\biggr)'~.
\label{eomflrw}
\end{equation}
The string energy per unit $\sigma$, in comoving units, is
$\epsilon\equiv\sqrt{{\bf x}^{'2}/(1-\dot{\bf x}^2)}$, implying that
the string energy is $\mu a \int \epsilon d\sigma$.  Equation
(\ref{eomflrw}) leads to
\begin{equation}
\frac{\dot\epsilon}{\epsilon}=-2\frac{\dot a}{a}\dot{\bf x}^2~.
\end{equation}
One usually fixes entirely the gauge by choosing $\sigma$ so that
$\epsilon=1$ initially.

\subsubsection{Cosmic Strings in Flat Spacetime}

In flat spacetime spacetime, the string equations of motion take the form
\begin{equation}
\partial_a(\sqrt{-\gamma}\gamma^{ab}x^\mu_{,b})=0~.
\end{equation}
We impose the conformal gauge
\begin{equation}
\dot x\cdot x'=0~ ~ ,~ ~ \dot x^2+x'^2=0~,
\label{cem}
\end{equation}
where overdots denote derivatives with respect to $\zeta^0$ and primes
denote derivatives with respect to $\zeta^1$.  In this gauge the
string equations of motion is just a two-dimensional wave equation,
\begin{equation}
\ddot{\bf x}-{\bf x}''=0~.
\label{eomm}
\end{equation}
To fix entirely the gauge, we also impose 
\begin{equation}
t\equiv x^0=\zeta^0~,
\end{equation}
which allows us to write the string trajectory as the three
dimensional vector ${\bf x}(\sigma,t)$, where $\zeta^1\equiv\sigma$,
the spacelike parameter along the string. This implies that the
constraint equations, Eq.~(\ref{cem}), and the string equations of
motion, Eq.~(\ref{eomm}), become
\begin{eqnarray}
\dot{\bf x}\cdot{\bf x}'&=&0\nonumber\\
\dot{{\bf x}}^2+{\bf x}'^2&=&1\nonumber\\
\ddot{\bf x}-{\bf x}''&=&0~.
\label{semf}
\end{eqnarray}
The above equations imply that the string moves perpendicularly to
itself with velocity $\dot{\bf x}$, that $\sigma$ is proportional to
the string energy, and that the string acceleration in the string rest
frame is inversely proportional to the local string curvature radius.
A curved string segment tends to straighten itself, resulting to
string oscillations.

The general solution to the string equation of motion in flat
spacetime, Eq.~(\ref{semf}c), is
\begin{equation}
{\bf x}=\frac{1}{2}\biggl[{\bf a}(\sigma-t)+{\bf
b}(\sigma+t)\biggr]~,
\end{equation}
where ${\bf a}(\sigma-t)$ and ${\bf b}(\sigma+t)$ are two
continuous arbitrary functions which satisfy
\begin{equation}
{\bf a}'^2={\bf b}'^2=1~.
\end{equation}
Thus, $\sigma$ is the length parameter along the three-dimensional
curves ${\bf a}(\sigma), {\bf b}(\sigma)$.

\subsubsection{Cosmic String Intercommutations}

The Goto-Nambu action describes to a good approximation cosmic string
segments which are separated. However, it leaves unanswered the issue
of what happens when strings cross. Numerical simulations have shown
that the ends of strings exchange partners, {\sl intercommute}, with
probability equal to 1.  These results have been confirmed for
global~\cite{nsig}, local~\cite{nsil}, and superconducting~\cite{nsis}
strings.

\begin{figure}[t]
\centerline{\includegraphics[width=3.4in]{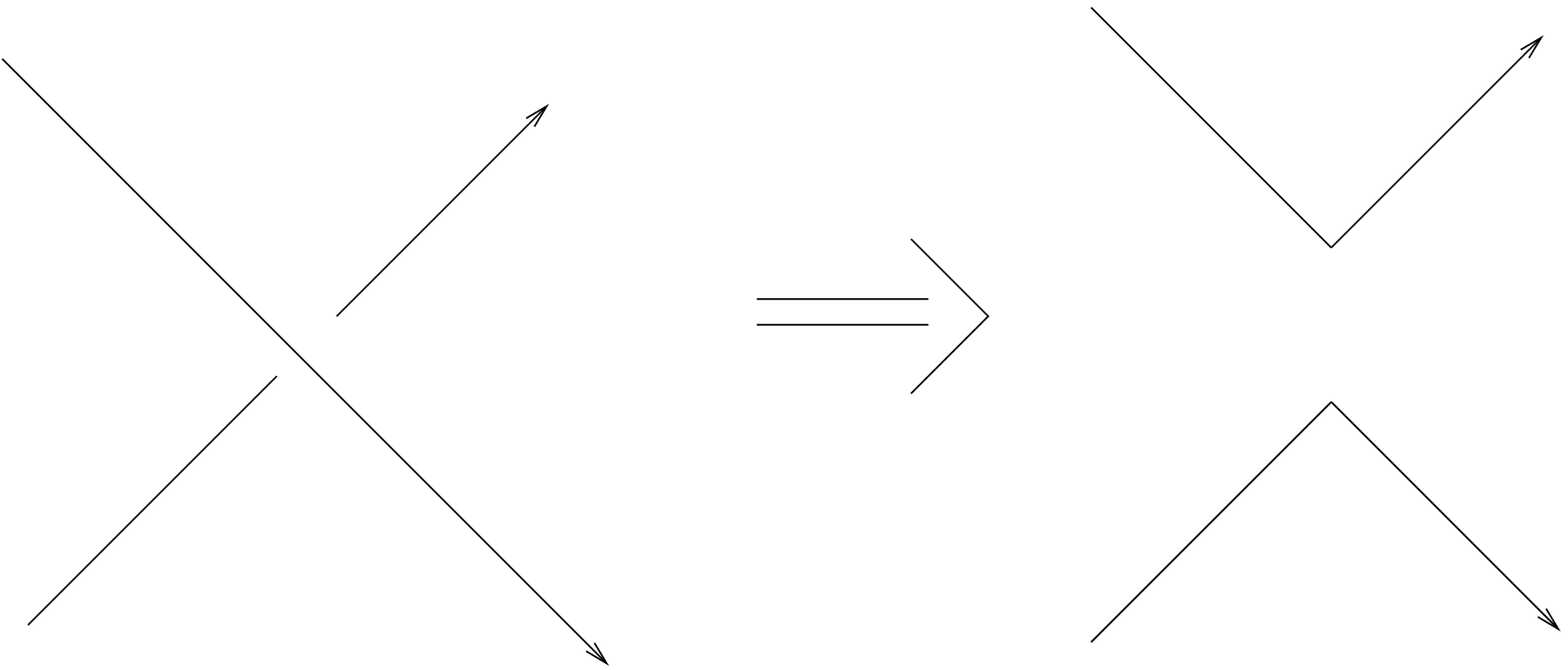}}
\vskip.2truecm
\centerline{\includegraphics[width=3.4in]{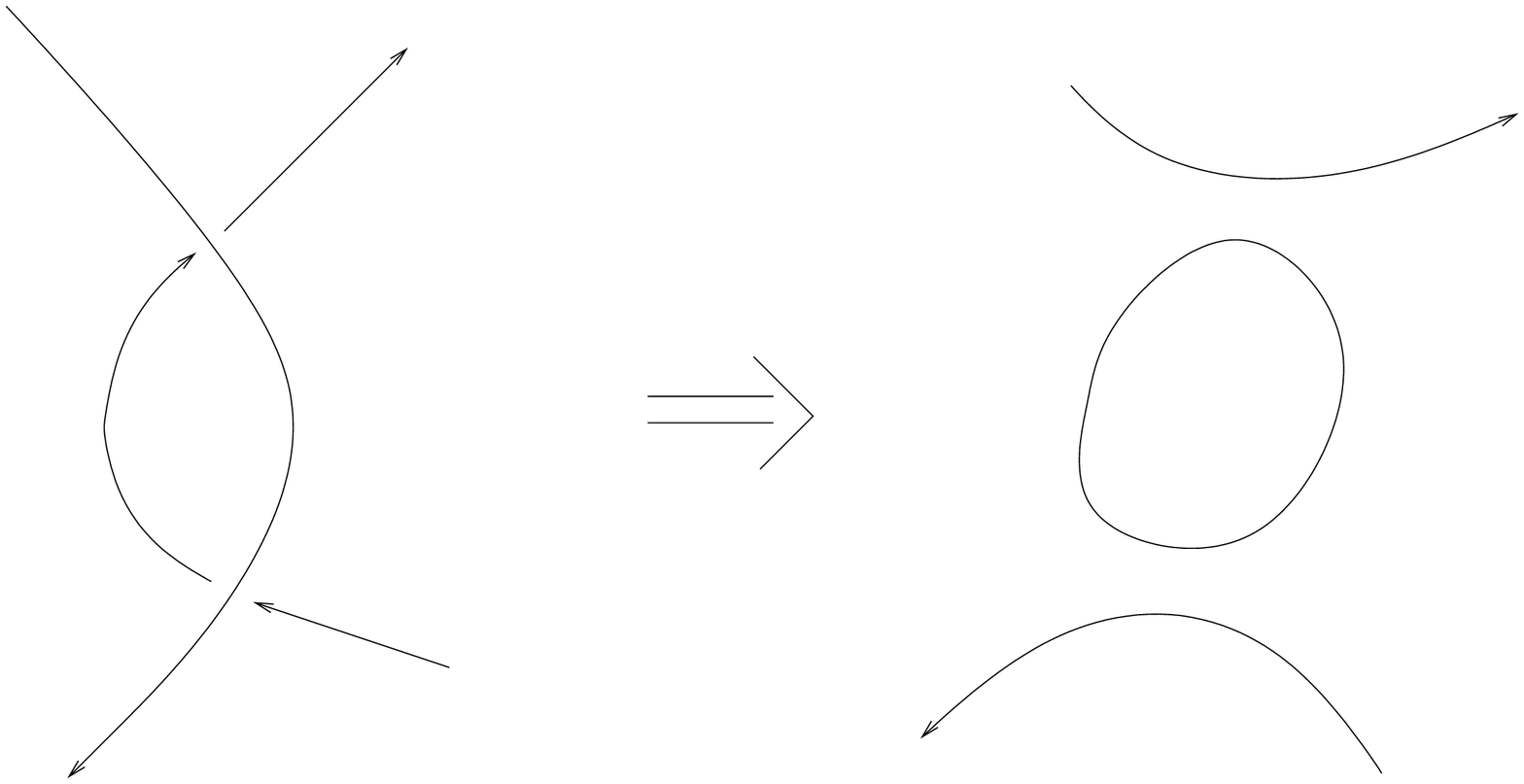}}
\vskip.2truecm
\centerline{\includegraphics[width=3.4in]{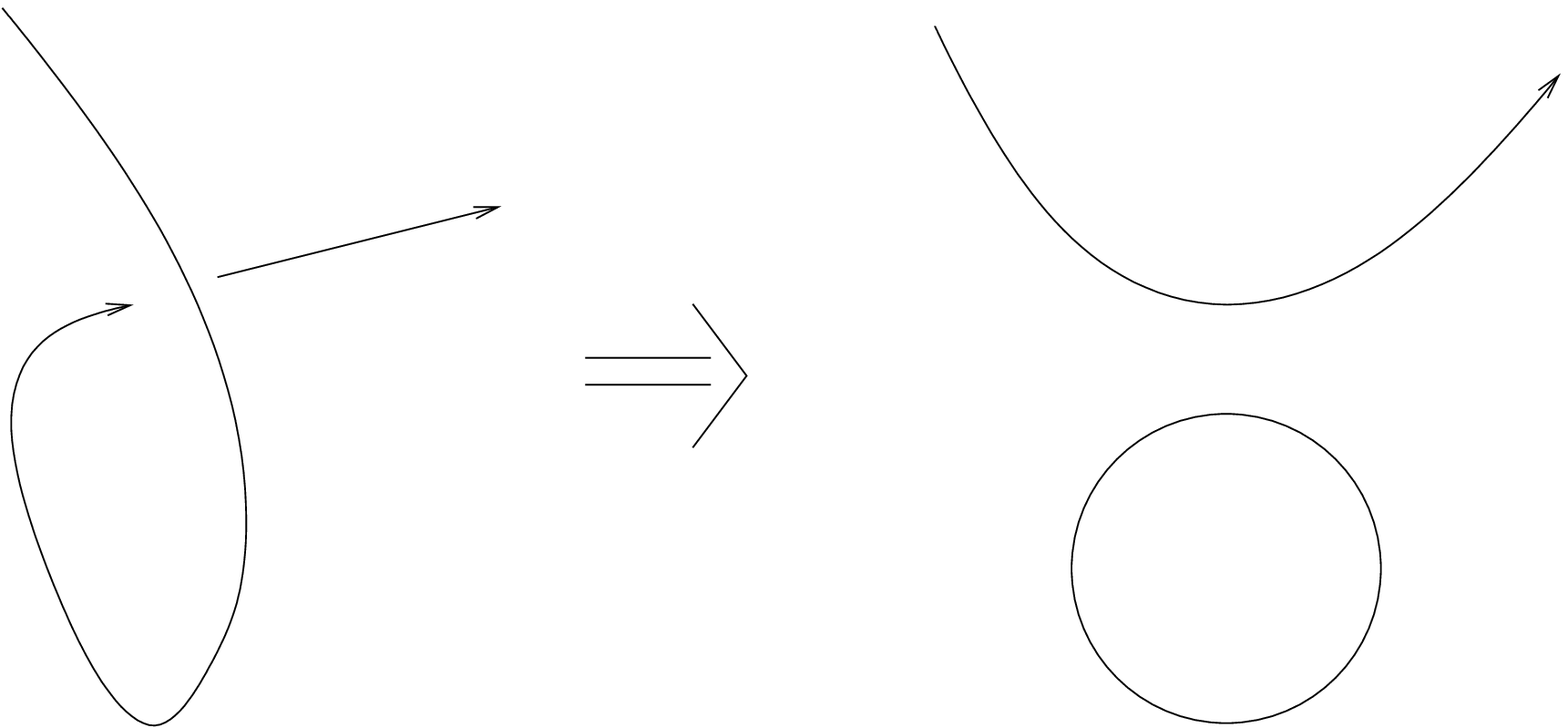}}
\caption{Illustration of string intersections: (a) string-string
intersection in one point, leading to the formation of two new long
strings via exchange of partners; (b) string-string intersections in
two points, leading to the formation of two new long strings via
exchange of partners, and one closed loop; and (c) self-string
intersections leading to the formation of one long string and a closed
string loop~\cite{msjcap}.}
\label{intersFig}
\end{figure}

String-string and self-string intersections leading to the formation
of new long strings and loops are drawn in Fig.~\ref{intersFig}.
Clearly string intercommutations produce discontinuities in $\dot{\bf
x}$ and ${\bf x}'$ on the new string segments at the intersection
point. These discontinuities, {\sl kinks}, are composed of right- and
left-moving pieces travelling along the string at the speed of light.

\subsection{Cosmic String Evolution}

Early analytic work~\cite{one-scale} identified the key property of
{\sl scaling}, where at least the basic properties of the string
network can be characterised by a single length scale, roughly the
persistence length or the interstring distance $\xi$, which grows with
the horizon.  This result was supported by subsequent numerical
work~\cite{numcs}.  However, further investigation revealed dynamical
processes, including loop production, at scales much smaller than
$\xi$~\cite{proc,mink}.

The cosmic string network can be divided into long (infinite) strings
and small loops. The energy density of long strings in the scaling
regime is given by (in the radiation era)
\begin{equation}
\rho_{\rm L}={\tilde \kappa}\mu t^{-2}~,
\end{equation}
where ${\tilde \kappa}$ is a numerical coefficient (${\tilde
\kappa}=20\pm 10$).  The small loops, their size distribution, and the
mechanism of their formation remained for years the least understood
parts of the string evolution.

Assuming that the long strings are characterised by a single length
scale $\xi(t)$, one gets 
\begin{equation}
\xi(t)=\biggl(\frac{\rho_{\rm L}}{\mu}\biggr)^{-1/2}={\tilde
\kappa}^{-1/2}t~.
\end{equation}
Thus, the typical distance between the nearest string segments and the
typical curvature radius of the strings are both of the order of
$\xi$. Early numerical simulations have shown that indeed the typical
curvature radius of long strings and the characteristic distance
between the strings are both comparable to the evolution time $t$.
Clearly, these results agree with the picture of the scale-invariant
evolution of the string network and with the one-scale hypothesis.

However, the numerical simulations have also shown~\cite{mink,exp1}
that small-scale processes (such as the production of small closed
loops) play an essential r\^ole in the energy balance of long strings.
The existence of an important small scale in the problem was also
indicated~\cite{mink} by an analysis of the string shapes.  In
response to these findings, a three-scale model was
developed~\cite{3-scale}, which describes the network in therms of
three scales, namely the usual energy density scale $\xi$, a
correlation length ${\bar\xi}$ along the string, and a scale $\zeta$
relating to local structure on the string. The small-scale structure
(wiggliness), which offers an explanation for the formation of the
small sized loops, is basically developed through intersections of long
string segments. It seemed likely from the three-scale model that
$\xi$ and ${\bar\xi}$ would scale, with $\zeta$ growing slowly, if at
all, until gravitational radiation effects became important when
$\zeta/\xi\approx 10^{-4}$~\cite{gr1,gr2}. Thus, according to the
three-scale model, the small length scale may reach scaling only if
one considers the gravitational back reaction effect.  Aspects of the
three-scale model have been checked~\cite{gmm} evolving a cosmic
string network is Minkowski spacetime. However, it was found that
loops are produced with tiny sizes, which led the authors to
suggest~\cite{gmm} that the dominant mode of energy loss of a cosmic
string network is particle production and not gravitational radiation
as the loops collapse almost immediately. One can find in the
literature studies which support~\cite{mark} this finding, and others
which they do not~\cite{aswa},\cite{msm}.

\begin{figure}[t]
\centerline{\includegraphics[width=3.4in]{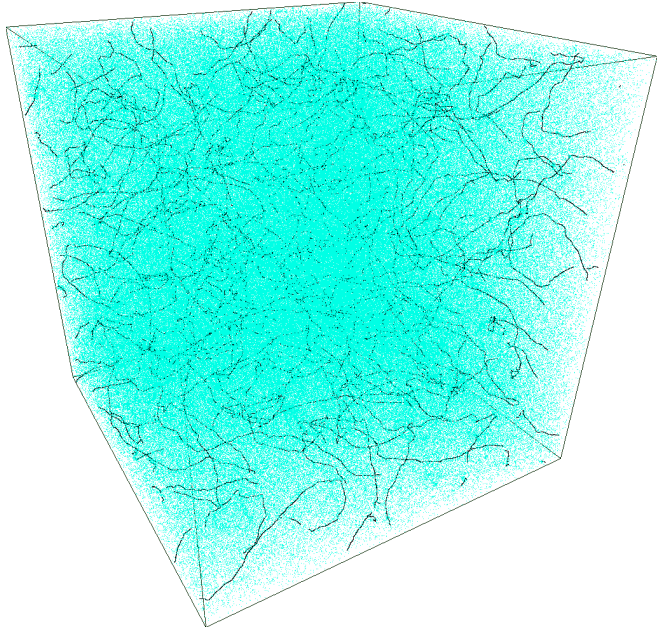}}
\caption{Snapshot of a network of long strings and closed loops in the
matter era~\cite{cmf}.}
\label{simulFig}
\end{figure}

Very recently, numerical simulations of cosmic string evolution in a
FLRW Universe ({\sl see}, Fig.~\ref{simulFig}), found
evidence~\cite{cmf} of a scaling regime for the cosmic string loops in
the radiation and matter dominated eras down to the hundredth of the
horizon time. It is important to note that the scaling was found
without considering any gravitational back reaction effect; it was
just the result of string intercmmuting mechanism. As it was reported
in Ref.~\cite{cmf}, the scaling regime of string loops appears after a
transient relaxation era, driven by a transient overproduction of
string loops with length close to the initial correlation length of
the string network. Calculating the amount of energy momentum tensor
lost from the string network, it was found~\cite{cmf} that a few
percents of the total string energy density disappear in the very
brief process of formation of numerically unresolved string loops
during the very first timesteps of the string evolution. Subsequently,
two other studies support these
findings~\cite{vov},\cite{martinsshellard}.

\subsection{String Thermodynamics}

It is one of the basic facts about string theory that the degeneracy
of string states increases exponentially with energy,
\begin{equation}
d(E)\sim e^{\beta_{\rm H}E}~.
\end{equation}
A consequence of this is that there is a maximum temperature $T_{\rm
max}=1/\beta_{\rm H}$, the Hagedorn
temperature~\cite{hagedorn,ed1,ed2}. In the microcanonical ensemble
the description of this situation is as follows: Consider a system of
closed string loops in a three-dimensional box.  Intersecting strings
intercommute, but otherwise they do not interact and are described by
the Goto-Nambu equations of motion. The statistical properties of a
system of strings in equilibrium are characterised by only one
parameter, the energy density of strings, $\rho$,
\begin{equation}
\rho=\frac{E}{L^3}~,
\end{equation}
where $L$ denotes the size of the cubical box.  The behaviour of the
system depends on whether it is at low or high energy densities, and
it undergoes a phase transition at a critical energy density, the
Hagedorn energy density $\rho_{\rm H}$.  Quantisation implies a lower
cutoff for the size of the string loops, determined by the string
tension $\mu$. The lower cutoff on the loop size is roughly
$\mu^{-1/2}$, implying that the mass of the smallest string loops is
$m_0\sim \mu^{1/2}$.

For a system of strings at the low energy density regime
($\rho\ll\rho_{\rm H}$), all strings are chopped down to the loops of
the smallest size, while larger loops are exponentially suppressed.
Thus, for small enough energy densities, the string equilibrium
configuration is dominated by the massless modes in the quantum
description.  The energy distribution of loops, given by the number
$dn$ of loops with energies between $E$ and $E+dE$ per unit volume,
is~\cite{ed1,ed2,ed3}
\begin{equation}
dn\propto e^{-\alpha E} E^{-5/2} dE ~~(\rho\ll\rho_{\rm H})~,
\end{equation}
where $\alpha=(5/2m_0)\ln(\rho_{\rm H}/\rho)$. 

However, as we increase the energy density, more and more oscillatory
modes of strings get excited. In particular, if we reach a critical
energy density, $\rho_{\rm H}$, then long oscillatory string states
begin to appear in the equilibrium state. The density at which this
happens corresponds to the Hagedorn temperature. The Hagedorn energy
density $\rho_{\rm H}$, achieved when the separation between the
smallest string loops is of the order of their sizes, is $\rho_{\rm
c}\sim m_0^4$. At the Hagedorn energy density the system undergoes a
phase transition characterised by the appearance of infinitely long
strings.  

At the high energy density regime ($\rho\gg\rho_{\rm c}$), the energy
distribution of string loops is~\cite{ed1,ed2,ed3}
\begin{equation}
dn= A m_0^{9/2} E^{-5/2} dE ~~(\rho\gg\rho_{\rm H})~,
\label{edlher}
\end{equation}
where $A$ is a numerical coefficient independent of $m_0$ and of
$\rho$. Equation (\ref{edlher}) implies that the mean-square radius
$R$ of the closed string loops is
\begin{equation}
R\sim m_0^{-3/2}E^{1/2}~,
\label{msr}
\end{equation}
meaning that large string loops are random walks of step $\sim
m_0^{-1}$.
Equations (\ref{edlher}) and (\ref{msr}) imply 
\begin{equation}
dn= A' R^{-4} dR ~~(\rho\gg\rho_{\rm H})~,
\label{sidclher}
\end{equation}
where $A'$ is a numerical constant. From Eq.~(\ref{sidclher}) one
concludes that the distribution of closed string loops is scale
invariant, since it does not depend on the cutoff parameter $m_0$.

The total energy density in finite string loops is independent of
$\rho$. Increasing the energy density $\rho$ of the system of strings,
the extra energy $E-E_{\rm H}$, where $E_{\rm H}=\rho_{\rm H}L^3$,
goes into the formation of infinitely long strings, implying
\begin{equation}
\rho-\rho_{\rm inf}={\rm const}  ~~(\rho\gg\rho_{\rm H})~,
\label{rhoinf}
\end{equation}
where $\rho_{\rm inf}$ denotes the energy density in infinitely long
strings. 

Clearly, the above analysis describes the behaviour of a system of
strings of low or high energy densities, while there is no analytic
description of the phase transition and of the intermediate densities
around the critical one, $\rho\sim\rho_{\rm H}$. An experimental
approach to the problem has been proposed in Ref.~\cite{smma1} and
later extended in Ref.~\cite{smed}.

The equilibrium properties of a system of cosmic strings have been
studied numerically in Ref.~\cite{smma1}. The strings are moving in a
three-dimensional flat space and the initial string states are chosen
to be a {\sl loop gas} consisting of the smallest two-point loops with
randomly assigned positions and velocities. This choice is made just
because it offers an easily adjustable string energy density. Clearly,
the equilibrium state is independent of the initial state. The
simulations revealed a distinct change of behaviour at a critical
energy density $\rho_{\rm H}=0.0172\pm 0.002$. For $\rho <\rho_{\rm
H}$, there are no infinitely long strings, thus their energy density,
$\rho_{\rm inf}$, is just zero. For $\rho >\rho_{\rm H}$, the energy
density in finite strings is constant, equal to $\rho_{\rm H}$, while
the extra energy goes to the infinitely long strings with energy
density $\rho_{\rm inf}=\rho-\rho_{\rm H}$.  Thus, Eqs.~(\ref{edlher})
and (\ref{rhoinf}) are valid for all $\rho>\rho_{\rm H}$, although
they were derived only in the limit $\rho\gg\rho_{\rm H}$.  At the
critical energy density, $\rho=\rho_{\rm H}$, the system of strings is
scale-invariant. At bigger energy densities, $\rho>\rho_{\rm H}$, the
energy distribution of closed string loops at different values of
$\rho$ were found ~\cite{smma1} to be identical within statistical
errors, and well defined by a line $dn/dE\propto E^{-5/2}$. Thus, for
$\rho>\rho_{\rm H}$, the distribution of finite strings is still
scale-invariant, but in addition the system includes the infinitely
long strings, which do not exhibit a scale-invariant distribution.
The number distribution for infinitely long strings goes as
$dn/dE\propto 1/E$, which means that the total number of infinitely
long strings is roughly $\log(E-E_{\rm H})$. So, typically the number of
long strings grows very slowly with energy; for $\rho>\rho_{\rm H}$
there are just a few infinitely long strings, which take up most of
the energy of the system.

The above numerical experiment has been extended~\cite{smed} for
strings moving in a higher dimensional box. The Hagedorn energy
density was found for strings moving in boxes of dimensionality
$d_{\rm B}=3,4,5$~\cite{smed}:
\begin{eqnarray}
\rho_{\rm H} &=&
\left\{ \begin{array}{lcl}0.172\pm 0.002  \ \ \mbox{for}\ \ & d_{\rm B}=3\\
0.062\pm 0.001  \ \ \mbox{for}\ \ & d_{\rm B}=4\\
0.031\pm 0.001  \ \ \mbox{for}\ \ & d_{\rm B}=5
\end{array}\right.
\end{eqnarray}
Moreover, the size distribution of closed finite string loops at the
high energy density regime was found to be independent of the
particular value of $\rho$ for a given dimensionality of the box
$d_{\rm B}$.  The size distribution of finite closed string loops was
found\cite{smed} to be well defined by a line
\begin{equation}
\frac{dn}{dE}\sim E^{-(1+d_{\rm B}/2)}~,
\end{equation}
where the space dimensionality $d_{\rm B}$ was taken equal to 3, 4, or
5 . The statistical errors indicated a slope equal to $-(1+d_{\rm
B}/2)\pm 0.2$. Above the Hagedorn energy density the system is again
characterised by a scale-invariant distribution of finite closed
string loops and a number of infinitely long strings with a
distribution which is not scale invariant.

\subsection{Genericity of Cosmic Strings Formation within SUSY GUTs}

The Standard Model (SM), even though it has been tested to a very high
precision, is incapable of explaining neutrino
masses~\cite{SK,SNO,kamland}.  An extension of the SM gauge group can
be realised within Supersymmetry (SUSY). SUSY offers a solution to the
gauge hierarchy problem, while in the supersymmetric extension of
the standard model the gauge coupling constants of the strong, weak
and electromagnetic interactions meet at a single point, $M_{\rm GUT}
\simeq (2-3) \times 10^{16}$ GeV.  In addition, SUSY GUTs provide the
scalar field which could drive inflation, explain the
matter-antimatter asymmetry of the Universe, and propose a candidate,
the lightest superparticle, for cold dark matter. We will address the
question of whether cosmic string formation is generic, in the context
of SUSY GUTs.

Within SUSY GUTs there is a large number of SSB patterns leading from
a large gauge group G to the SM gauge group G$_{\rm SM}\equiv$
SU(3)$_{\rm C}\times$ SU(2)$_{\rm L}\times$ U(1)$_{\rm Y}$. The study
of the homotopy group of the false vacuum for each SSB scheme
determines whether there is defect formation and identifies the type
of the formed defect. Clearly, if there is formation of domain walls
or monopoles, one will have to place an era of supersymmetric hybrid
inflation to dilute them. To consider a SSB scheme as a successful
one, it should be able to explain the matter/anti-matter asymmetry of
the Universe and to account for the proton lifetime
measurements~\cite{SK}.  

In what follows, we consider a mechanism of baryogenesis via
leptogenesis, which can be thermal or nonthermal one.  In the case of
nonthermal leptogenesis, U(1)$_{\rm B-L}$ (B and L, are the baryon and
lepton numbers, respectively) is a sub-group of the GUT gauge group,
G$_{\rm GUT}$, and B-L is broken at the end or after inflation. If one
considers a mechanism of thermal leptogenesis, B-L is broken
independently of inflation. If leptogenesis is thermal and B-L is
broken before the inflationary era, then one should check whether the
temperature at which B-L is broken --- this temperature defines the
mass of the right-handed neutrinos --- is smaller than the reheating
temperature. To have a successful inflationary cosmology, the
reheating temperature should be lower than the limit imposed by the
gravitino. To ensure the stability of proton, the discrete symmetry
Z$_2$, which is contained in U(1)$_{\rm B-L}$, must be kept unbroken
down to low energies. Thus, the successful SSB schemes should end at
G$_{\rm SM}\times$ Z$_2$. Taking all these considerations into account
we will examine within all acceptable SSB patterns, how often cosmic
strings form at the end of the inflationary era.

To proceed, one has to first choose the large gauge group G$_{\rm
GUT}$.  In Ref.~\cite{jrs} this study has been done in detail for a
large number of simple Lie groups. Considering GUTs based on simple
gauge groups, the type of supersymmetric hybrid inflation will be of
the F-type. The minimum rank of G$_{\rm GUT}$ has to be at least equal
to 4, to contain the G$_{\rm SM}$ as a subgroup.  Then one has to
study the possible embeddings of G$_{\rm SM}$ in G$_{\rm GUT}$ so that
there is an agreement with the SM phenomenology and especially with
the hypercharges of the known particles. Moreover, the large gauge
group G$_{\rm GUT}$ must include a complex representation, needed to
describe the SM fermions, and it must be anomaly free.  In principle,
${\rm SU}(n)$ may not be anomaly free. We thus assume that all ${\rm
SU}(n)$ groups we consider have indeed a fermionic representation
which certifies that the model is anomaly free. We set as the upper
bound on the rank $r$ of the group, $r\leq 8$. Clearly, the choice of
the maximum rank is in principle arbitrary.  This choice could, in a
sense, be motivated by the Horava-Witten~\cite{hw} model, based on
${\rm E}_8\times {\rm E}_8$.  Concluding, the large gauge group
G$_{\rm GUT}$ could be one of the following: SO(10), E$_6$, SO(14),
SU(8), SU(9); flipped SU(5) and [SU(3)]$^3$ are included within this
list as subgroups of SO(10) and E$_6$, respectively.

A detailed study of all SSB schemes which bring us from G$_{\rm GUT}$
down to the SM gauge group G$_{\rm SM}$, by one or more intermediate
steps, shows that cosmic strings are generically formed at the end of
hybrid inflation.  If the large gauge group G$_{\rm GUT}$ is SO(10)
then cosmic strings formation is unavoidable~\cite{jrs}.  

The genericity of cosmic string formation for ${\rm E}_6$ depends
whether one considers thermal or nonthermal leptogenesis. More
precisely, under the assumption of nonthermal leptogenesis, cosmic
string formation is unavoidable. Considering thermal leptogenesis,
cosmic strings formation at the end of hybrid inflation arises in
$98\%$ of the acceptable SSB schemes~\cite{adp}.  Finally,
if the requirement of having Z$_2$ unbroken down to low energies is
relaxed and thermal leptogenesis is considered as the mechanism for
baryogenesis, cosmic string formation accompanies hybrid inflation in
$80\%$ of the SSB schemes.  

The SSB schemes of SU(6) and SU(7) down to the G$_{\rm SM}$ which
could accommodate an inflationary era with no defect (of any kind) at
later times are inconsistent with proton lifetime
measurements. Minimal SU(6) and SU(7) do not predict neutrino
masses~\cite{jrs}, implying that these models are incompatible with
high energy physics phenomenology. 

Higher rank groups, namely SO(14), SU(8) and SU(9), should in general
lead to cosmic string formation at the end of hybrid inflation. In all
these schemes, cosmic string formation is sometimes accompanied by the
formation of embedded strings. The strings which form at the end of
hybrid inflation have a mass which is proportional to the inflationary
scale.

\section{Cosmic Microwave Background Temperature Anisotropies}

The CMB temperature anisotropies offer a powerful test for theoretical
models aiming at describing the early Universe.  The characteristics
of the CMB multipole moments can be used to discriminate among
theoretical models and to constrain the parameters space.

The spherical harmonic expansion of the CMB temperature anisotropies,
as a function of angular position, is given by
\begin{equation}
\label{dTT}
\frac{\delta T}{T}({\bf n})=\sum _{\ell m}a_{\ell m} {\cal W}_\ell
Y_{\ell m}({\bf n})~\,
\ \ \ \mbox {with}\ \ \ 
a_{\ell m}=\int {\rm
d}\Omega _{{\bf n}}\frac{\delta T}{T}({\bf n})Y_{\ell m}^*({\bf n})~;
\end{equation}
${\cal W}_\ell $ stands for the $\ell$-dependent window function of
the particular experiment.  The angular power spectrum of CMB
temperature anisotropies is expressed in terms of the dimensionless
coefficients $C_\ell$, which appear in the expansion of the angular
correlation function in terms of the Legendre polynomials $P_\ell$:
\begin{equation}
\biggl \langle 0\biggl |\frac{\delta T}{T}({\bf n})\frac{\delta T}{
T}({\bf n}') \biggr |0\biggr\rangle \left|_{{~}_{\!\!({\bf n\cdot
n}'=\cos\vartheta)}}\right. = \frac{1}{4\pi}\sum_\ell(2\ell+1)C_\ell
P_\ell(\cos\vartheta) {\cal W}_\ell^2 ~,
\label{dtovertvs}
\end{equation}
where we have used the addition theorem of spherical harmonics, \ie,
\begin{equation}
\sum _{m=-\ell}^\ell Y_{\ell m}({\bf n})Y_{\ell m}^\star({\bf n}')
=\frac{2\ell+1}{4\pi}P_\ell({\bf n}\cdot{\bf n}')~.
\end{equation}
It compares points in the sky separated by an angle $\vartheta$.
Here, the brackets denote spatial average, or expectation values if
perturbations are quantised. Equation (\ref{dtovertvs}) holds only if
the initial state for cosmological perturbations of quantum-mechanical
origin is the vacuum~\cite{jam,gms}.  The value of $C_\ell$ is
determined by fluctuations on angular scales of the order of
$\pi/\ell$. The angular power spectrum of anisotropies observed today
is usually given by the power per logarithmic interval in $\ell$,
plotting $\ell(\ell+1)C_\ell$ versus $\ell$.

To find the power spectrum induced by topological defects, one has to
solve, in Fourier space for each given wave vector ${\bf k}$ a
system of linear perturbation equations with random sources:
\begin{equation}
{\cal D} X = {\cal S}~,
\label{deq} 
\end{equation}
where ${\cal D}$ denotes a time dependent linear differential
operator, $X$ is a vector which contains the various matter
perturbation variables, and ${\cal S}$ is the random source term,
consisting of linear combinations of the energy momentum tensor of the
defect.  For given initial conditions, Eq.~(\ref{deq}) can be solved
by means of a Green's function, ${\cal G}(\tau,\tau ')$, in the form
\begin{equation}
X_j(\tau_0,{\bf k}) =\int_{\tau_{in}}^{\tau_0}\! d\tau{\cal G}_{jm}
(\tau_0,\tau,{\bf k})
{\cal S}_m(\tau,{\bf k})~.
\label{Gfunctsol}
\end{equation}
To compute power spectra or, more generally, quadratic expectation
values of the form $\langle X_j(\tau_0,{\bf k})X_m^*(\tau_0,{\bf
k'})\rangle$, one has to calculate
\begin{eqnarray}
&&\langle X_j(\tau_0,{\bf k})X_l^\star(\tau_0,{\bf k}')\rangle =
\nonumber \\ && 
\ \ \ \ \ \ \ \ \int_{\tau_{in}}^{\tau_0}\! d\tau{\cal
G}_{jm}(\tau,{\bf k}) \int_{\tau_{in}}^{\tau_0} \! d\tau '{\cal
G}^\star_{ln}(\tau ',{\bf k}')\times \langle{\cal S}_m(\tau,{\bf
k}){\cal S}_n^\star(\tau ',{\bf k}')\rangle~.
\label{power}
\end{eqnarray} 
Thus, to compute power spectra, one should know the unequal time
two-point correlators $\langle{\cal S}_m(\tau,{\bf k}){\cal
S}_n^\star(\tau ',{\bf k}')\rangle$ in Fourier space~\cite{Hindarsh}.
This object is calculated by means of heavy numerical simulations.

The CMB temperature anisotropies provide a powerful tool to
discriminate among inflation and topological defects.  On large
angular scales ($\ell \lesssim 50$), both families of models lead to
approximately scale-invariant spectra, with however a different
prediction regarding the statistics of the induced
perturbations. Provided the quantum fields are initially placed in the
vacuum, inflation predicts generically Gaussian fluctuations, whereas
in the case of topological defect models, the induced perturbations
are clearly nongaussian, at least at sufficiently high angular
resolution. This is an interesting fingerprint, even though difficult
to test through the data. In the context of inflation, nongaussianity
can however also be present, as for example in the case of stochastic
inflation~\cite{stochastic}, or in a class of inflationary models
involving two scalar fields leading to nongaussian isothermal
fluctuations with a blue spectrum~\cite{linde1}. In addition, allowing
nonvacuum initial states for the cosmological perturbations of
quantum-mechanical origin, one generically obtains a non-Gaussian
spectrum~\cite{jam,gms}, in the context of single-field inflation.

On intermediate and small angular scales however, the predictions of
inflation are quite different than those of topological defect models,
due to the different nature of the induced perturbations.  On the one
hand, the inflationary fluctuations are coherent, in the sense that
the perturbations are initially at the same phase and subsequently
evolve linearly and independently of each other. The subsequent
progressive phase shift between different modes produces the {\sl
acoustic peak} structure. On the other hand, in topological defect
models, fluctuations are constantly induced by the sources
(defects). Since topological defects evolve in a nonlinear manner, and
since the random initial conditions of the source term in the
perturbation equations of a given scale leaks into other scales,
perfect coherence is destroyed.  The predictions of the defects models
regarding the characteristics of the CMB spectrum are:
\begin{itemize}
\item
Global ${\cal O}(4)$ textures lead to
a position of the first acoustic peak  at $\ell\simeq 350$
with an amplitude $\sim 1.5$ times higher than the Sachs-Wolfe
plateau~\cite{rm}.
\item
Global ${\cal O}(N)$ textures in the
large $N$ limit lead to a quite flat spectrum, with a slow decay after
$\ell \sim 100$~\cite{dkm}. Similar are the predictions of other
global ${\cal O}(N)$ defects \cite{clstrings,num}.
\item
Local cosmic strings simulations~\cite{csmark} found a broad peak at
$\ell\approx 150-400$, being produced from both vector and scalar
modes, which peaks at $\ell\approx 180$ and $\ell\approx 400$
respcetively.

\end{itemize}
The position and amplitude of the acoustic peaks, as found by the CMB
measurements~\cite{maxi,boom,dasi,wmap}, are in disagreement with the
predictions of topological defect models. As a consequence, CMB
measurements rule out pure topological defect models as the origin of
initial density perturbations leading to the observed structure
formation. 

\subsection{Mixed Models}

\begin{figure}[t]
\vskip.8truecm
\centerline{\includegraphics[width=2.2in,angle=270]{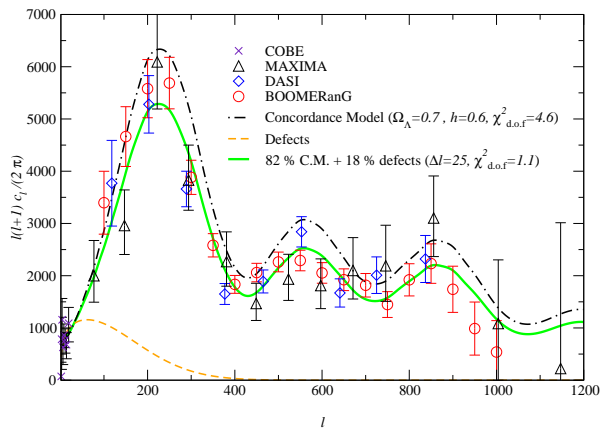}}
\vskip.4truecm
\caption{$\ell (\ell + 1) C_\ell$ versus $\ell$ for three different
    models. The string
    contribution turns out to be $\sim 18 \%$ of the total~\cite{bprs}.}
\label{csCMBFig}
\end{figure}

Since cosmic strings are expected to be generically formed in the
context of SUSY GUTs, one should consider {\sl mixed perturbation
models} where the dominant r\^ole is played by the inflaton field but cosmic
strings have also a contribution, small but not negligible.  
Restricting ourselves to the angular power spectrum, we can remain in the
linear regime. In this case, 
\begin{equation}
C_\ell =   \alpha     C^{\scriptscriptstyle{\rm I}}_\ell
         + (1-\alpha) C^{\scriptscriptstyle{\rm S}}_\ell~,
\label{cl}
\end{equation}
where $C^{\scriptscriptstyle{\rm I}}_\ell$ and $C^{\scriptscriptstyle
{\rm S}}_\ell$ denote the (COBE normalized) Legendre coefficients due
to adiabatic inflaton fluctuations and those stemming from the cosmic
string network, respectively. The coefficient $\alpha$ in
Eq.~(\ref{cl}) is a free parameter giving the relative amplitude for
the two contributions.  Comparing the $C_\ell$, calculated using
Eq.~(\ref{cl}) --- where $C^{\scriptscriptstyle{\rm I}}_\ell$ is taken
from a generic inflationary model and $C^{\scriptscriptstyle {\rm
S}}_\ell$ from numerical simulations of cosmic string networks ---
with data obtained from the most recent CMB measurements, one gets
that a cosmic string contribution to the primordial fluctuations
higher than $14\%$ is excluded up to $95\%$ confidence
level~\cite{bprs,pogosian,wyman} ({\sl see}, Fig.~\ref{csCMBFig}).

In what follows, we follow a conservative approach and do not allow
cosmic strings to contribute more than $10\%$ to the CMB temperature
anisotropies.
 
\subsection{Supersymmetric Hybrid Inflation}

Inflation offers simple answers to the shortcomings of the standard
hot big bang model. In addition, simple inflationary models offer
successful candidates for the initial density fluctuations leading to
the observed structure formation.  

One crucial question though is to answer how generic is the onset of
inflation~\cite{ecms,gt,us} and to find consistent and natural models
of inflation from the point of view of particle physics.  Even though
one can argue that the initial conditions which favor inflationary
models are the likely outcome of the quantum era before
inflation~\cite{ecms}, one should then show that inflation will last
long enough to solve the shortcomings of the standard hot big bang
model~\cite{gt,us}. In addition, to find natural ways to guarantee the
flatness of the inflaton potential remains a difficult task. Inflation
is, unfortunately, still a paradigm in search of a model.  It is thus
crucial to identify successful but natural inflationary models
motivated from high energy physics.

In what follows we discuss two well-studied inflationary models in the
framework of supersymmetry, namely F/D-term inflation. Our aim is to
check the compatibility of these models --- here cosmic string
inflation is generic --- with the CMB and gravitino constraints.

\subsubsection{F-term Inflation}

F-term inflation can be naturally accommodated in the framework of
GUTs, when a GUT gauge group G$_{\rm GUT}$ is broken down to the
G$_{\rm SM}$ at an energy $M_{\rm GUT}$, according to the scheme
\begin{equation}
{\rm G}_{\rm GUT} \stackrel{M_{\rm GUT}}{\hbox to 0.8cm
{\rightarrowfill}} {\rm H}_1 \xrightarrow{9}{M_{\rm
infl}}{1}{\Phi_+\Phi_-} {\rm H}_2 {\longrightarrow} {\rm G}_{\rm SM}~;
\end{equation}
$\Phi_+, \Phi_-$ is a pair of GUT Higgs superfields in nontrivial
complex conjugate representations, which lower the rank of the group
by one unit when acquiring nonzero vacuum expectation value. The
inflationary phase takes place at the beginning of the symmetry
breaking ${\rm H}_1\stackrel{M_{\rm infl}}{\longrightarrow} {\rm
H}_2$.

F-term inflation is based on the globally supersymmetric
renormalisable superpotential
\begin{equation}\label{superpot}
W_{\rm infl}^{\rm F}=\kappa  S(\Phi_+\Phi_- - M^2)~,
\end{equation}
where $S$ is a GUT gauge singlet left handed superfield, $\Phi_+$ and
$\Phi_-$ are defined above; $\kappa$ and $M$ are two constants ($M$
has dimensions of mass) which can be taken positive with field
redefinition.  The chiral superfields $S, \Phi_+, \Phi_-$ are
taken to have canonical kinetic terms.  This superpotential is the
most general one consistent with an R-symmetry under which $W
\rightarrow e^{i \beta} W~, \Phi_- \rightarrow e^{-i \beta} \Phi_-~,
\Phi_+ \rightarrow e^{i \beta} \Phi_+$, and $ S \rightarrow e^{i
\beta} S$. An R-symmetry can ensure that the rest of the
renormalisable terms are either absent or irrelevant.

The scalar potential reads
\begin{equation}
\label{scalpot1}
V(\phi_+,\phi_-, S)= |F_{\Phi_+}|^2+|F_{\Phi_-}|^2+|F_ S|^2
+\frac{1}{2}\sum_a g_a^2 D_a^2~.
\end{equation}
The F-term is such that $F_{\Phi_i} \equiv |\partial W/\partial
\Phi_i|_{\theta=0}$, where we take the scalar component of the
superfields once we differentiate with respect to $\Phi_i=\Phi_+,
\Phi_-,  S$. The D-terms are
\begin{equation}
D_a=\bar{\phi}_i\,{(T_a)^i}_j\,\phi^j +\xi_a~,
\end{equation}
with $a$ the label of the gauge group generators $T_a$, $g_a$ the
gauge coupling, and $\xi_a$ the Fayet-Iliopoulos term. By definition,
in the F-term inflation the real constant $\xi_a$ is zero; it can only
be nonzero if $T_a$ generates an extra U(1) group.  In the context of
F-term hybrid inflation, the F-terms give rise to the inflationary
potential energy density, while the D-terms are flat along the
inflationary trajectory, thus one may neglect them during inflation.

The potential has one valley of local minima, $V=\kappa^2 M^4$, for
$S> M $ with $\phi_+ = \phi_-=0$, and one global supersymmetric
minimum, $V=0$, at $S=0$ and $\phi_+ = \phi_- = M$. Imposing initially
$ S \gg M$, the fields quickly settle down the valley of local
minima.  Since in the slow roll inflationary valley, the ground state
of the scalar potential is nonzero, SUSY is broken.  In the tree
level, along the inflationary valley the potential is constant,
therefore perfectly flat. A slope along the potential can be generated
by including the one-loop radiative corrections. Thus, the scalar
potential gets a little tilt which helps the inflaton field $S$ to
slowly roll down the valley of minima. The one-loop radiative
corrections to the scalar potential along the inflationary valley,
lead to an effective potential~\cite{DvaShaScha,Lazarides,SenoSha,rs1}
\begin{eqnarray}
\label{VexactF}
V_{\rm eff}^{\rm F}(|S|)&&=\kappa^2M^4\biggl\{1+\frac{\kappa^2
\cal{N}}{32\pi^2}\biggl[2\ln\frac{|S|^2\kappa^2}{\Lambda^2}\nonumber\\
&&+\biggl(\frac{|S|^2}{ M^2}+1\biggr)^2\ln\big(1+\frac{M^2}{|S|^2})
+\biggl(\frac{|S|^2}{M^2}-1\biggr)^2
\ln\biggl(1-\frac{M^2}{|S|^2}\biggr)\biggr]\biggr\}
\end{eqnarray}
where $\Lambda$ is a renormalisation scale and $\cal{N}$ stands for
the dimensionality of the representation to which the complex scalar
components $\phi_+, \phi_-$ of the chiral superfields $\Phi_+, \Phi_-$
belong. For example, ${\cal N}={\bf 27, 126, 351}$, correspond to
realistic SSB schemes in SO(10), or E$_6$ models.

Considering only large angular scales, \ie, taking only the Sachs-Wolfe
contribution, one can get the contributions to the CMB temperature
ani\-so\-tro\-pies analytically. The quadrupole anisotropy has one
contribution coming from the inflaton field, splitted into scalar and
tensor modes, and one contribution coming from the cosmic string
network, given by numerical simulations~\cite{ls}.  The inflaton field
contribution is
\begin{equation}
\biggl(\frac{\delta T}{T}\biggr)_{\rm Q-infl}=
\biggl[\biggl(\frac{\delta T}{T}\biggr)^2_{\rm Q-scal}+
\biggl(\frac{\delta T}{T}\biggr)^2_{\rm Q-tens}\biggr]^{1/2}~,
\label{qic}
\end{equation}
where the quadrupole anisotropy due to the scalar and tensor
Sachs-Wolfe effect is
\begin{eqnarray}
\biggl(\frac{\delta T}{T}\biggr)_{\rm Q-scal}&=&
\frac{1}{4\sqrt{45}\pi}\frac{V^{3/2}(\varphi_{\rm Q})}
{M^3_{\rm Pl}V'(\varphi_{\rm Q})}\nonumber\\
\biggl(\frac{\delta T}{T}\biggr)_{\rm Q-tens}&\sim&
\frac{0.77}{8\pi}\frac{V^{1/2}(\varphi_{\rm Q})}{M^2_{\rm Pl}}~,
\label{scal-tens}
\end{eqnarray}
respectively, with $V'\equiv dV(\varphi)/d\varphi$, $M_{\rm Pl}$ the
reduced Planck mass, $M_{\rm Pl}=(8\pi G)^{-1/2}\simeq 2.43\times
10^{18}{\rm GeV}$, and $\varphi_{\rm Q}$ the value of the inflaton
field when the comoving scale corresponds to the quadrupole anisotropy
became bigger than the Hubble radius.  It can be calculated using
Eqs.~(\ref{VexactF}),(\ref{qic}),(\ref{scal-tens}).

Fixing the number of e-foldings to 60, the inflaton and cosmic
string contribution to the CMB, for a given gauge group G$_{\rm GUT}$,
depend on the superpotential coupling $\kappa$, or equivalently on
the symmetry breaking scale $M$ associated with the inflaton mass
scale, which coincides with the string mass scale. The relation
between $\kappa$ and $M$ is
\begin{equation}
\label{Mdekappa}
\frac{M}{M_{\rm Pl}}=\frac{\sqrt{N_{\rm Q} \cal N}\,\kappa}{2\,
\pi\,y_{\rm Q}}~,
\end{equation}
where 
\begin{equation}
y_{\rm Q}^2=\int_1^{\frac{|S_{\rm Q}|^2}{M^2}} dz\biggl[z\big\{
(z+1)\ln(1+z^{-1})+(z-1)\ln(1-z^{-1})\bigl\}\big]~.
\end{equation}
The total quadrupole anisotropy has to be normalised to the COBE data.

\begin{figure}[hhh]
\begin{center}
\includegraphics[scale=.6]{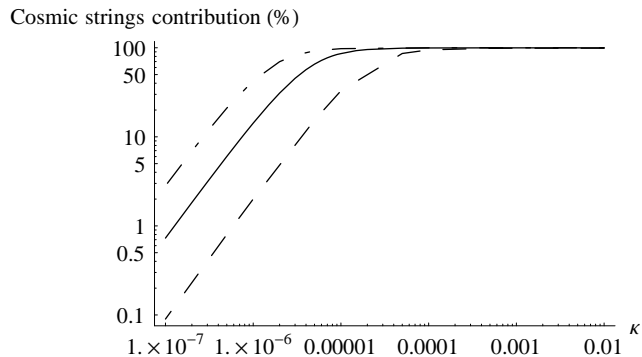}
\caption{Contribution of cosmic strings to the quadrupole anisotropy
as a function of the superpotential coupling $\kappa$. The three
curves correspond to $\mathcal{N}=\mathbf{27}$ (curve with broken
line), $\mathcal{N}= \mathbf{126}$ (full line) and
$\mathcal{N}=\mathbf{351}$ (curve with lines and
dots)~\cite{rs1}.}\label{contribCSplot}
\end{center}
\end{figure}

A detailed study has been performed in Ref.~\cite{rs1}. It was shown
that the cosmic string contribution is consistent with the CMB
measurements, provided~\cite{rs1}
\begin{equation}
M\lsim 2\times 10^{15} {\rm GeV} ~~\Leftrightarrow ~~\kappa \lsim
7\times10^{-7}~.
\label{constrkappa}
\end{equation}
In Fig.\ref{contribCSplot}, one can see the contribution of cosmic
strings to the quadrupole anisotropy as a function of the  superpotential 
coupling $\kappa$~\cite{rs1}. The three curves correspond
to $\mathcal{N}=\mathbf{27}$ (curve with broken line), $\mathcal{N}=
\mathbf{126}$ (full line) and $\mathcal{N}=\mathbf{351}$ (curve with
lines and dots).

The constraint on $\kappa$ given in Eq.~(\ref{constrkappa}) is in
agreement with the one found in Ref.~\cite{lk}.  Strictly speaking the
above condition was found in the context of SO(10) gauge group, but
the conditions imposed in the context of other gauge groups are of the
same order of magnitude since $M$ is a slowly varying function of the
dimensionality ${\cal N}$ of the representations to which the scalar
components of the chiral Higgs superfields belong.

The superpotential coupling $\kappa$ is also subject to the gravitino
constraint which imposes an upper limit to the reheating temperature,
to avoid gravitino overproduction.  The reheating temperature $T_{\rm
RH}$ characterises the reheating process via which the Universe enters
the high entropy radiation dominated phase at the end of the
inflationary era.  Within the minimal supersymmetric standard model
and assuming a see-saw mechanism to give rise to massive neutrinos,
the reheating temperature is~\cite{rs1}
\begin{equation}
T_{\rm RH}\approx \frac{(8\pi)^{1/4}}{7}(\Gamma M_{\rm Pl})^{1/2}~,
\label{rhn}
\end{equation}
where $\Gamma$ is the decay width of the oscillating inflaton and
Higgs fields into right-handed neutrinos
\begin{equation}
\Gamma=\frac{1}{8\pi}\biggl(\frac{M_i}{M}\biggr)^2m_{\rm infl}~;
\label{g}
\end{equation}
with $m_{\rm infl}=\sqrt 2 \kappa M$ the inflaton mass and $M_i$ the
right-handed neutrino mass eigenvalue with $M_i<m_{\rm infl}/2$.
Equations (\ref{Mdekappa}),~(\ref{rhn}),~(\ref{g}) lead to
 \begin{equation}
 T_{\rm RH}\sim \frac{1}{12}\left(\frac{60}{ N_{\rm Q}}\right)^{1/4}
 \left(\frac{1}{{ \cal N}}\right)^{1/4}    y_{\rm
 Q}^{1/2} M_i~.
 \end{equation}
In order to have successful reheating, it is important not to create
too many gravitinos, which imply the following constraint on the
reheating temperature~\cite{gr} $T_{\rm RH}\leq 10^9 {\rm GeV}$.
Since the two heaviest neutrinos are expected to have masses of the
order of $M_3\simeq 10^{15}$ GeV and $M_2\simeq 2.5 \times 10^{12}$
GeV respectively~\cite{pati}, $M_i$ is identified with $M_1\sim 6\times
10^9$ GeV~\cite{pati}. The gravitino constraint on $\kappa$
reads~\cite{rs1} $\kappa \lsim 8\times 10^{-3}$, which is clearly a
weaker constraint than the one imposed from the CMB data. 

Concluding, F-term inflation leads generically to cosmic string
formation at the end of the inflationary era. The cosmic strings
formed are of the GUT scale. This class of models can be compatible
with CMB measurements, provided the superpotential coupling is smaller
\footnote{The linear mass density $\mu$ gets a correction due to
deviations from the Bogomol'nyi limit, which may enlare~\cite{enlarge}
the parameter space for F-term inflation. Note that this does not hold for
D-term inflation, since then strings are BPS (Bogomol'nyi-Prasad-Sommerfield) 
 states.}
 than $10^{-6}$ .  This tuning of the free parameter
$\kappa$ can be softened if one allows for the curvaton mechanism.

\begin{figure}[t]
\centerline{\includegraphics[width=3.4in]{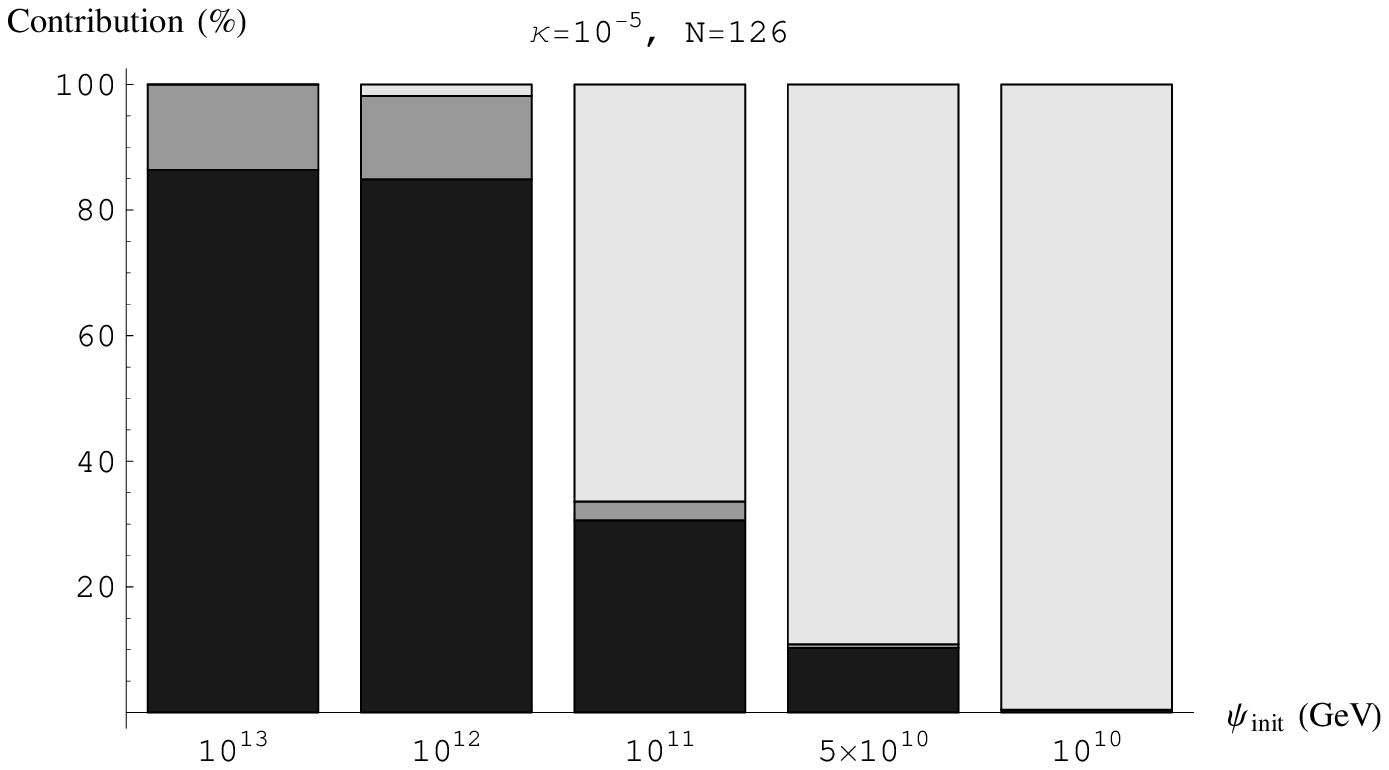}}
\vskip.2truecm
\centerline{\includegraphics[width=3.4in]{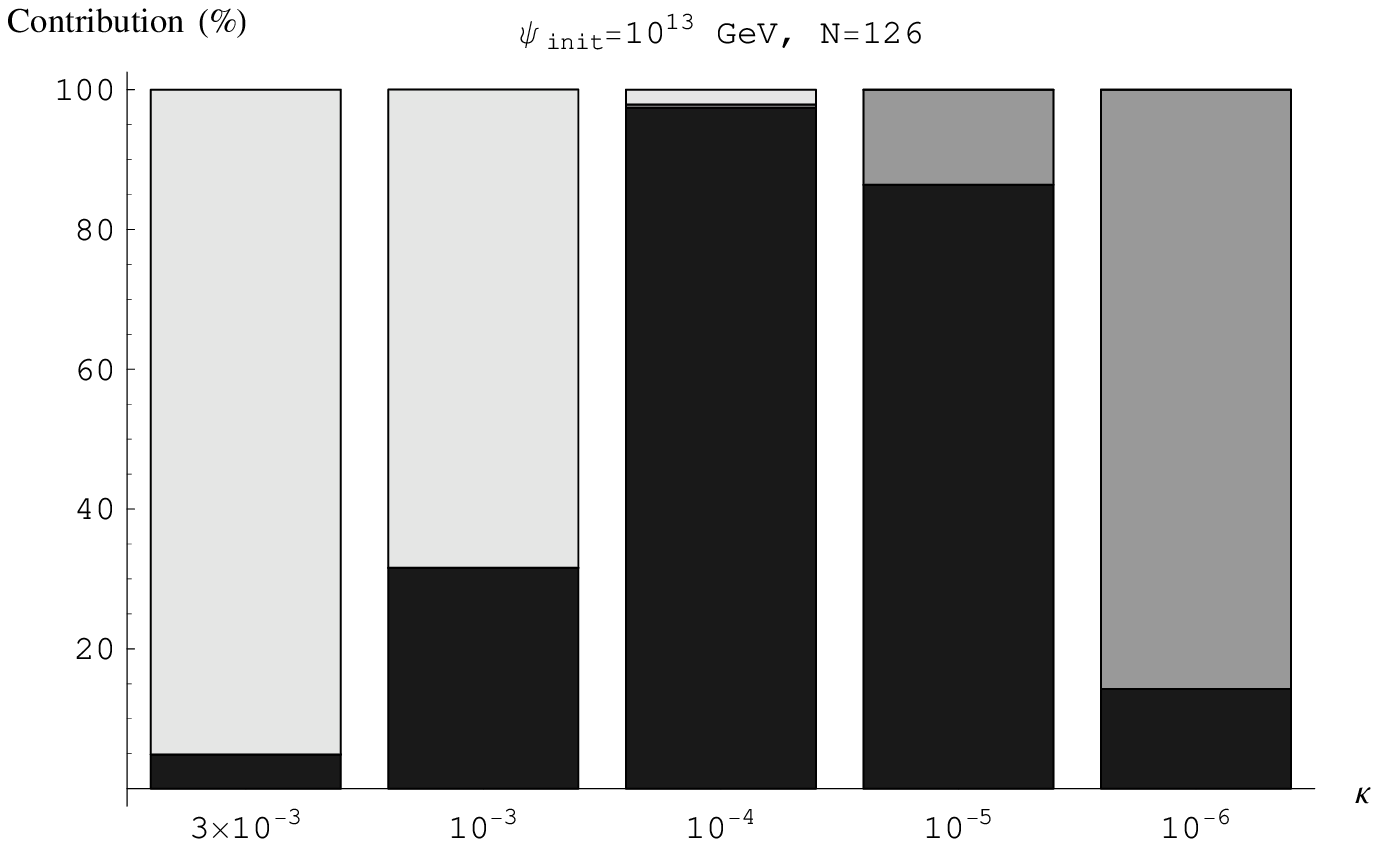}}
\caption{The cosmic strings (dark gray), curvaton (light gray) and
inflaton (gray) contributions to the CMB temperature anisotropies as a
function of the the initial value of the curvaton field
${\cal\psi}_{\rm init}$, and the superpotential coupling 
$\kappa$, for ${\cal N}=\mathbf{126}$~\cite{rs1}.}
\label{contribcurvFig}
\end{figure}

According to the curvaton mechanism~\cite{lw2002,mt2001}, another
scalar field, called the curvaton, could generate the initial density
perturbations whereas the inflaton field is only responsible for the
dynamics of the Universe. The curvaton is a scalar field, that is
sub-dominant during the inflationary era as well as at the beginning
of the radiation dominated era which follows the inflationary
phase. There is no correlation between the primordial fluctuations of
the inflaton and curvaton fields. Clearly, within supersymmetric
theories such scalar fields are expected to exist. In addition,
embedded strings, if they accompany the formation of cosmic strings,
they may offer a natural curvaton candidate, provided the decay
product of embedded strings gives rise to a scalar field before the
onset of inflation.  Considering the curvaton scenario, the coupling
$\kappa$ is only constrained by the gravitino limit. More precisely,
assuming the existence of a curvaton field, there is an additional
contribution to the temperature anisotropies. The WMAP CMB
measurements impose~\cite{rs1} the following limit on the initial
value of the curvaton field
\begin{equation}
{\cal\psi}_{\rm init} \lsim 5\times 10^{13}\,\left( 
\frac{\kappa}{10^{-2}}\right){\rm GeV}~,
\end{equation}
provided the parameter $\kappa$ is in the range $[10^{-6},~1]$ 
({\sl see}, Fig.~\ref{contribcurvFig}).

The above results hold also if one includes supergravity corrections.
This is expected since the value of the inflaton field is several
orders of magnitude below the Planck scale.

\subsubsection{D-term Inflation}

The early history of the Universe at energies below the Planck scale
is described by an effective N=1 supergravity (SUGRA)
theory. Inflation should have taken place at an energy scale
$V^{1/4}\lesssim 4\times 10^{16}$ GeV, implying that inflationary
models should be constructed in the framework of SUGRA. 

However, it is difficult to implement slow-roll inflation within
SUGRA. The positive false vacuum of the inflaton field breaks
spontaneously global supersymmetry, which gets restored after the end
of inflation. In supergravity theories, the supersymmetry breaking is
transmitted to all fields by gravity, and thus any scalar field,
including the inflaton, gets an effective mass of the order of the the
expansion rate $H$ during inflation. 

This problem, known as the problem of {\sl Hubble-induced mass},
originates from F-term interactions --- note that it is absent in the
model we have described in the previous subsection --- and thus it is
resolved if one considers the vacuum energy as being dominated by
non-zero D-terms of some superfields~\cite{dterm1,dterm2}. This result
led to a dramatic interest in D-term inflation, since in addition, it
can be easily implemented within string theory.

D-term inflation is derived from the superpotential
\begin{equation}
\label{superpotD}
W^{\rm D}_{\rm infl}=\lambda S \Phi_+\Phi_-~;
\end{equation}
$S, \Phi_-, \Phi_+$ are three chiral superfields and $\lambda$ is the
superpotential coupling.  D-term inflation requires the existence of a
nonzero Fayet-Iliopoulos term $\xi$, which can be added to the
Lagrangian only in the presence of an extra U(1) gauge symmetry, under
which, the three chiral superfields have charges $Q_S=0$,
$Q_{\Phi_+}=+1$, and $Q_{\Phi_-}=-1$. This extra U(1) gauge symmetry
can be of a different origin; hereafter we consider a nonanomalous
U(1) gauge symmetry.  Thus, D-term inflation requires a scheme, like
\begin{equation}
{\rm G}_{\rm GUT}\times {\rm U}(1) \stackrel{M_{\rm GUT}}{\hbox to
0.8cm{\rightarrowfill}} {\rm H} \times {\rm U}(1) \xrightarrow{9}{M_{\rm
nfl}}{1}{\Phi_+\Phi_-} {\rm H} \rightarrow {\rm G}_{\rm SM}~.
\end{equation}
The symmetry breaking at the end of the inflationary phase implies
that cosmic strings are always formed at the end of D-term hybrid
inflation. To avoid cosmic strings, several mechanisms have been
proposed which either consider more complicated models or require
additional ingredients. For example, one can add a nonrenormalisable
term in the potential~\cite{shifted}, or add an additional discrete
symmetry~\cite{smooth}, or consider GUT models based on nonsimple
groups~\cite{mcg}, or introduce a new pair of charged
superfields~\cite{jaa} so that cosmic string formation is avoided at
the end of D-term inflation. In what follows, we show that standard
D-term inflation followed unavoidably by cosmic string production is
compatible with CMB data, because the cosmic string contribution to
the CMB data is not constant nor dominant. Thus, one does not have to
invoke some new physics.

In the global supersymmetric limit,
Eqs.~(\ref{scalpot1}),~(\ref{superpotD}) lead to the following
expression for the scalar potential
\begin{equation}
\label{VtotD}
V^{\rm D}(\phi_+,\phi_-,S) = \lambda^2
\left[\,|S|^2(|\phi_+|^2+|\phi_-|^2) + |\phi_+\phi_-|^2 \right]
+\frac{g^2}{2}(|\phi_+|^2-|\phi_-|^2+\xi)^2~,
\end{equation}
where $g$ is the gauge coupling of the U(1) symmetry and $\xi$ is a
Fayet-Iliopoulos term, chosen to be positive.

In D-term inflation, as opposed to F-term inflation, the inflaton mass
acquires values of the order of Planck mass, and therefore, the
correct analysis must be done in the framework of SUGRA. The SSB of
SUSY in the inflationary valley introduces a splitting in the masses
of the components of the chiral superfields $\Phi_\pm$. As a result,
we obtain~\cite{rs2} two scalars with squared masses
$m^2_{\pm}=\lambda^2|S|^2 \exp\left(|S|^2/M^2_{\rm Pl}\right)\pm g^2
\xi$ and a Dirac fermion with squared mass $m_{\rm f}^2=\lambda^2|S|^2
\exp\left(|S|^2/M^2_{\rm Pl}\right)$.  Calculating the radiative
corrections, the effective scalar potential for minimal supergravity
reads~\cite{rs1,rs2}
\begin{eqnarray}
\label{vDsugra}
V_{\rm eff} 
=\frac{g^2\xi^2}{2}\biggl\{1+\frac{g^2}{16 \pi^2}
&&\times\biggl[2\ln\frac{|S|^2\lambda^2}{\Lambda^2}e^{\frac{|S|^2}{
M^2_{\rm Pl}}}\nonumber\\
+&&\biggl( \frac{\lambda^2 |S|^2}{
g^2\xi}e^{\frac{|S|^2}{M_{\rm Pl}^2}} +1\biggr)^2
\ln\biggl(1+\frac{ g^2\xi}{\lambda^2 |S|^2 }e^{-\frac{|S|^2}{ M_{\rm Pl}^2}}
\biggr)\nonumber\\
+&&\biggl( \frac{\lambda^2 |S|^2}{
g^2\xi}e^{\frac{|S|^2}{ M_{\rm Pl}^2}} -1\biggr)^2
\ln\biggl(1-\frac{ g^2\xi}{\lambda^2 |S|^2 }e^{-\frac{|S|^2}{ M_{\rm Pl}^2}}
\biggr)\biggr]\biggr\}
\end{eqnarray}

\begin{figure}[t]
\centerline{\includegraphics[width=4.4in]{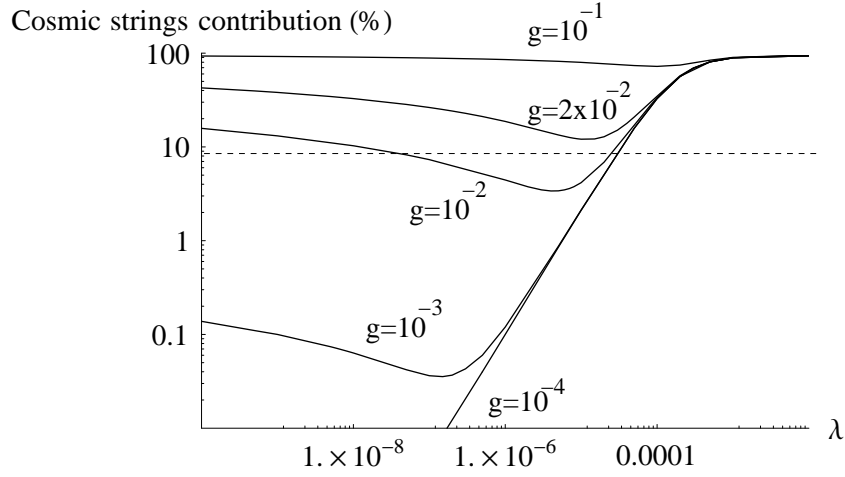}}
\caption{Cosmic string contribution to the CMB temperature
anisotropies as a function of the superpotential coupling $\lambda$
for different values of the gauge coupling $g$. The maximal
contribution allowed by WMAP is represented by a dotted
line~\cite{rs1,rs2}.}
\label{dtermFig}
\end{figure}

As it was explicitely shown in Refs.~\cite{rs1,rs2}, D-term inflation
can be compatible with current CMB measurements; the cosmic strings
contribution to the CMB is model-dependent. The results obtained in
Refs.~\cite{rs1,rs2} can be summarised as follows: (i) $g\gsim 2\times
10^{-2}$ is incompatible with the allowed cosmic string contribution
to the WMAP measurements; (ii) for $g\lsim 2\times 10^{-2}$ the
constraint on the superpotential coupling $\lambda$ reads $\lambda
\lsim 3\times 10^{-5}$; (iii) SUGRA corrections impose in addition a
lower limit to $\lambda$; (iv) the constraints induced on the
couplings by the CMB measurements can be expressed as a single
constraint on the Fayet-Iliopoulos term $\xi$, namely $\sqrt\xi \lsim
2\times 10^{15}~{\rm GeV}$. They are shown in Fig.~[\ref{dtermFig}].

\begin{figure}[t]
\centerline{\includegraphics[width=3.4in]{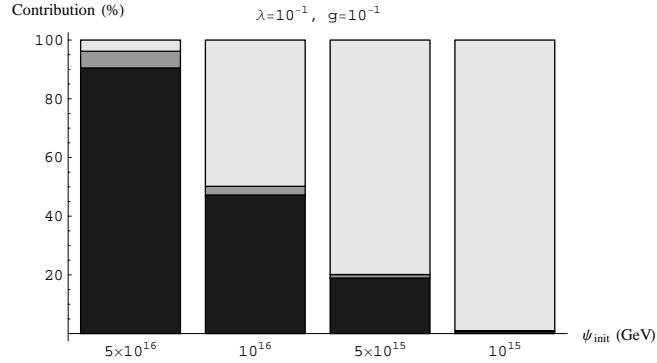}}
\caption{The cosmic strings (dark gray), curvaton (light gray) and
inflaton (gray) contributions to the CMB temperature anisotropies as a
function of the the initial value of the curvaton field
${\cal\psi}_{\rm init}$, for $\lambda=10^{-1}$ and
$g=10^{-1}$~\cite{rs2}.}
\label{dtermconstrcurvFig}
\end{figure}

Assuming the existence of a curvaton field, the fine tuning on the
couplings can be avoided provided~\cite{rs1,rs2}
\begin{equation}
\psi_{\rm init}\lsim 3\times 10^{14}\left(\frac{g}{ 10^{-2}}\right) ~{\rm GeV}
~~~{\rm for}~~ \lambda\in [10^{-1}, 10^{-4}]~.
\end{equation}
Clearly, for smaller values of $\lambda$, the curvaton mechanism is
not necessary. We show in Fig.~\ref{dtermconstrcurvFig} the three
contributions as a function of $\psi_{\rm init}$, for
$\lambda=10^{-1}$ and $g=10^{-1}$.  There are values of
$\psi_{\rm init}$ which allow bigger values of the superpotential
coupling $\lambda$ and of the gauge coupling $g$, than the upper
bounds obtained in the absence of a curvaton field.

Concluding, standard D-term inflation always leads to cosmic string
formation at the end of the inflationary era; these cosmic strings are
of the grand unification scale. This class of models is still
compatible with CMB measurements, provided the couplings are small
enough.  As in the case of F-term inflation the fine tuning of the
couplings can be softened provided one considers the curvaton
mechanism. In this case, the imposed CMB constraint on the initial
value of the curvaton field reads~\cite{rs1,rs2}
\begin{equation}
\psi_{\rm init}\lsim 3\times 10^{14}\left(\frac{g}{ 10^{-2}}\right)
~{\rm GeV}~, ~~\mbox{for}~~\lambda\in [10^{-1}, 10^{-4}]~.
\end{equation}

The above conclusions are still valid in the
revised version of D-term inflation, in the framework of SUGRA with
constant Fayet-Iliopoulos terms. In the context of N=1, 3+1 SUGRA, the
presence of constant Fayet-Iliopoulos terms shows up in covariant
derivatives of all fermions. In addition, since the relevant local
U(1) symmetry is a gauged R-symmetry~\cite{toine2}, the constant
Fayet-Iliopoulos terms also show up in the supersymmetry
transformation laws.  In Ref.~\cite{toine1} there were presented all
corrections of order $g\xi/M_{\rm Pl}^2$ to the classical SUGRA action
required by local supersymmetry.  Under U(1) gauge transformations in
the directions in which there are constant Fayet-Iliopoulos terms
$\xi$, the superpotential must transform as~\cite{toine2}
\begin{equation}
\delta W=-i\frac{g\xi}{ M_{\rm Pl}^2}W~,
\end{equation}
otherwise the constant Fayet-Iliopoulos term $\xi$ vanishes. This
requirement is consistent with the fact that in the gauge theory at
$M_{\rm Pl}\rightarrow \infty$ the potential is U(1) invariant. To
promote the simple SUSY D-term inflation model, Eq.~(\ref{superpotD}),
to SUGRA with constant Fayet-Iliopoulos terms, one has to change the
charge assignments for the chiral superfields, so that the
superpotential transforms under local R-symmetry~\cite{toine1}. In
SUSY, the D-term potential is neutral under U(1) symmetry, while in
SUGRA the total charge of $\Phi_\pm$ fields does not vanish but is
equal to $-\xi/M_{\rm Pl}^2$. More precisely, the D-term contribution
to the scalar potential $V$ [see Eq.~(\ref{VtotD})], should be
replaced by $(g^2/2)(q_+|\phi_+|^2+q_-|\phi_-|^2+\xi)^2$ where
\begin{equation}
q_\pm=\pm 1-\rho_\pm\frac{\xi}{ M_{\rm Pl}^2}\ \ \ \ \mbox {with} \ \ \
\ \rho_++\rho_-=1~.
\end{equation}
In addition, the squared masses of the scalar components $\phi_\pm$
become 
\begin{equation}
m^2_{\pm}=\lambda^2|S|^2 \exp\left(|S|^2/M^2_{\rm Pl}\right)\pm g^2
\xi q_\pm~;
\end{equation}
the Dirac fermion mass remains unchanged. 

For the limits we imposed on the Fayet-Iliopoulos term $\xi$, the
correction $\xi/M_{\rm Pl}^2$ is $\sim 10^{-6}$, implying that the
constraints we obtained on $g$ and $\lambda$, or equivalently on
$\sqrt{\xi}$, as well as the constraint on $\psi_{\rm init}$ still hold
in the revised version of D-term inflation within SUGRA~\cite{adp}.

It is important to generalise the above study in the case of
nonminimal SUGRA~\cite{inprep}, in order to know whether qualitatively
the above picture remains valid. A recent study~\cite{inprep} has
shown that non-minimal K\"ahler potential do not avoid the fine
tuning, since the cosmic string contribution remains dominant unless
the couplings and mass scales are small.  For example, taking into
account higher order corrections in the K\"ahler potential, or
considering supergravity with shift symmetry, we have
obtained~\cite{inprep} that the $9\%$ constraint in the allowed
contribution of cosmic strings in the CMB spectrum implies
\begin{equation}
\sqrt\xi\leq 2\times 10^{15} {\rm GeV}\ \Leftrightarrow \
G\mu\leq 8\times 10^{-7}~.
\end{equation}
The {\sl cosmic string problem} can be definitely cured if
one considers more complicated models, for example where strings
become topologically unstable, namely semi-local strings.

\section{Cosmic Superstrings}

At first, and for many years, cosmic strings and superstrings were
considered as two well separated issues. The main reason for this
clear distinction may be considered the Planckian tension of
superstrings. If the string mass scale is of the order of the Planck
mass, then the four-dimensional F- and D-string self gravity is
$G_4\mu_{\rm F}\sim {\cal O}(g_{\rm s}^2)$ and $G_4\mu_{{\rm D}1}\sim
{\cal O}(g_{\rm s})$ ($g_{\rm s}$ stands for the string coupling),
respectively, while current CMB measurements impose an upper limit on
the self gravity of strings of $G\mu< 10^{-6}$. Moreover, heavy
superstrings could have only been produced before inflation, and
therefore diluted. In addition, Witten showed~\cite{witten} that, in
the context of the heterotic theory, long fundamental BPS strings are
unstable, thus they would not survive on cosmic time scales; non-BPS
strings were also believed to be unstable.

At present, the picture has been dramatically changed (for a review,
see {\sl e.g.}, Ref.~\cite{maj}).  In the framework of braneworld
cosmology, our Universe represents a three-dimensional Dirichlet brane
(D3-brane) on which open fundamental strings (F-strings)
end~\cite{Polch}. Such a D3-brane is embedded in a higher dimensional
space, the bulk. Brane interactions can unwind and evaporate higher
dimensional branes, leaving behind D3-branes embedded in a higher
dimensional bulk; one of these D3-branes could play the r\^ole of our
Universe~\cite{dks}. Large extra dimensions can be employed to address
the hierarchy problem~\cite{Ark}, a result which lead to an increasing
interest in braneworld scenarios.  As it has been argued~\cite{jst,st}
D-brane-antibrane inflation leads to the production of
lower-dimensional D-branes, that are one-dimensional (D-strings) in
the noncompact directions.  The production of zero- and
two-dimensional defects (monopoles and domain walls, respectively) is
suppressed. The large compact dimensions and the large warp factors
can allow for superstrings of much lower tensions, in the range
between $10^{-11}< G\mu < 10^{-6}$.  Depending on the model of string
theory inflation, one can identify~\cite{cmp} D-strings, F-strings,
bound states of $p$ fundamental strings and $q$ D-strings for
relatively prime $(p,q)$, or no strings at all.

The probability that two colliding superstrings reconnect can be much
less than one. Thus, a reconnection probability ${\cal P}< 1$ is one of
the distinguishing features of superstrings.  D-strings can miss each
other in the compact dimension, leading to a smaller ${\cal P}$, while
for F-strings the scattering has to be calculated quantum mechanically,
since these are quantum mechanical objects.  

The collisions between all possible pairs of superstrings have been
studied in string perturbation theory~\cite{jjp}. For F-strings, the
reconnection probability is of the order of $g_{\rm s}^2$.  For F-F
string collisions, it was found~\cite{jjp} that the reconnection
probability $\cal P$ is $10^{-3}\lsim {\cal P}\lsim 1$. For D-D string
collisions, $10^{-1}\lsim{\cal P}\lsim 1$. Finally, for F-D string
collisions, the reconnection probability can take any value between 0
and 1.  These results have been confirmed~\cite{hh1} by a quantum
calculation of the reconnection probability for colliding D-strings.
Similarly, the string self-intersection probability is reduced. When
D- and F-strings meet they can form a three-string junction, with a
composite DF-string. In IIB string theory, they may be found bound
$(p,q)$ states of $p$ F-strings and $q$ D-strings, where $p$ and $q$
are coprime. This leads to the question of whether there are frozen
networks dominating the matter content of the Universe, or whether
scaling solutions can be achieved.

The evolution of cosmic superstring networks has been addressed
numerically~\cite{csn1,csn2,csn3,csn4,csn5,csn6} and
analytically~\cite{csanalyt}.

The first numerical approach~\cite{csn1}, studies independent
stochastic networks of D- and F-strings, evolving in a flat spacetime.
One can either evolve strings in a higher dimensional space keeping
the reconnection probability equal to 1, or evolve them in a
three-dimensional space with ${\cal P}\ll 1$. These two approaches
lead to results which are equivalent qualitatively, as it has been
shown in Ref.~\cite{csn1}.  These numerical simulations have shown
that the characteristic length scale $\xi$, giving the typical
distance between the nearest string segments and the typical
curvature of strings, grows linearly with time
\begin{equation}
 \xi(t)\propto \zeta t ~;
\end{equation}
the slope $\zeta$ depends on the reconnection probability ${\cal
P}$, and on the energy of the smallest allowed loops (i.e., the energy
cutoff).  For reconnection (or intercommuting) probability in the
range $10^{-3}\lsim {\cal P} \lsim 0.3$, it was found~\cite{csn1}
\begin{equation}
\zeta \propto \sqrt{\cal P} \Rightarrow \xi(t)\propto \sqrt{\cal P} t~,
\label{law}
\end{equation}
in agreement with older results~\cite{mink}. 

One can find in the literature statements claiming that $\xi(t)$
should be proportional to ${\cal P} t$ instead.  If this were correct,
then the energy density of cosmic superstrings of a given tension could
be considerably higher than that of their field theory analogues
(cosmic strings).  In Ref.~\cite{jst1} it is claimed that the energy
density of longs strings $\rho_{\rm l}$ evolves as $\dot\rho_{\rm
l}=2(\dot a/a)\rho_{\rm l}-{\cal P}(\rho_{\rm l}/\xi)$, where $H=\dot
a/a$ is the Hubble constant.  Substituting the ansatz
$\xi(t)=\gamma(t)t$, the authors of Ref.~\cite{jst1} obtain
$\dot\gamma=-[1/(2t)](\gamma-{\cal P})$, during the
radiation-dominated era. This equation has a stable fixed point at
$\gamma(t)={\cal P}$, implying that~\cite{jst1} $\xi\simeq {\cal P}
t$. However, Ref.~\cite{jst1} misses out the fact that intersections
between two long strings is not the most efficient mechanism for
energy loss of the string network. The possible string intersections
can be divided into three possible cases (see, Fig.~\ref{intersFig}):
(i) two long strings collide in one point and exchange partners with
intercommuting probability ${\cal P}_1$; (ii) two strings collide in
two points and exchange partners chopping off a small loop with
intercommuting probability ${\cal P}_1^2$; and (iii) one long string
self-intersects in one point and chops off a loop with intercommuting
probability ${\cal P}_2$, which in general is different than ${\cal
P}_1$.  Only cases (ii) and (iii) lead to a closed loop formation and
therefore remove energy from the long string network.  Between cases
(ii) and (iii), only case (iii) is an efficient way of forming loops
and therefore dissipating energy: case (iii) is more frequent than
case (ii), and case (ii) has in general a smaller probability, since
${\cal P}_1\sim {\cal P}_2$~\cite{csn1}.  However, the heuristic
argument employed in Ref.\cite{jst1} does not refer to self-string
intersections (i.e, case (iii)); it only applies to intersections
between two long strings, which depend on the string velocity.
However self-string intersections should not depend on how fast the
string moves, a string can intersect itself even if it does not move
but it just oscillates locally.

The findings of Ref.~\cite{csn1} cleared the misconception about the
behaviour of the scale $\xi$, and shown that the cosmic superstring energy
density may be higher than the field theory case, but at most only by
one order of magnitude. 

An important question to be addressed is whether cosmic superstrings
can survive for a long time and eventually dominate the energy density
of the Universe. This could lead to an overdense Universe with
catastrophic cosmological consequences. If the reconnection
probability is too low, or equivalently, the strings move in a higher
dimensional space and therefore miss each other even if ${\cal P}$ is
high, then one may fear that the string network does not reach a
scaling regime. The string energy density redshifts as $1/a^2$, where
$a$ stands for the scale factor. Since for a string network with
correlation length $\xi$, there is about 1 long string per horizon
volume, the string energy density is $\sim \mu/a^2$. String
interactions leading to loop formation guarantee a scaling regime, in
the sense that strings remain a constant fraction of the energy
density of the Universe. Loops do not feel the expansion of the
Universe, so they are not conformally stretched and they redshift as
$1/a^3$. As the loops oscillate, they lose their energy and they
eventually collapse. Clearly, scaling is not a trivial issue for cosmic
superstrings. 

In the first numerical approach~\cite{csn1}, where they have been only
considered independent stochastic networks of either F- or D-strings,
it was shown that each such network reaches a scaling regime. This has
been shown by either evolving strings in a higher dimensional space
with intercommuting probability equal to 1, or evolving strings in a
three-dimensional space with intercommuting probability much smaller
than 1. 

In a realistic case however, $(p,q)$ strings come in very large number
of different types, while a $(p,q)$ string can decay to a loop only if
it self-intersects of collide with another $(p,q)$ or $(-p,-q)$
string. A collision between $(p,q)$ and $(p',q')$ strings will lead to
a new string $(p\pm p',q\pm q')$, provided the end points of the
initial two strings are not attached to other three-string vertices,
thus they are not a part of a web.  If the collision between two
strings can lead to the formation of one new string, on a timescale
much shorter than the typical collision timescale, then the creation
of a web may be avoided, and the resulting network is composed by
strings which are on the average nonintersecting. Then one can imagine
the following configuration: A string network, composed by different
types of $(p,q)$ strings undergoes collisions and
self-intersections. Energy considerations imply the production of
lighter daughter strings, leading eventually to one of the following
strings: $(\pm 1,0), (0,\pm 1), \pm(1,1), \pm(1,-1)$. These ones may
then self-intersect, form loops and scale individually. Provided the
relative contribution of each of these strings to the energy density
of the Universe is small enough, the Universe will not be overclosed.

 This result has been confirmed by studying numerically the behavior
of a network of interacting Dirichlet-fundamental strings $(p,q)$ in
Ref.~\cite{csn3}. To model $(p,q)$ strings arising from
compactifications of type IIB string theory, the authors
studied~\cite{csn3} the evolution of nonabelian string networks. The
positive element of such nonabelian networks is that they contain
multiple vertices where many different types of string join together.
Such networks have the potential of leading to a string dominated
Universe due to tangled networks of interacting $(p,q)$ strings that
freeze.  It was shown~\cite{csn3} that such freezing does not take
place and the network reaches a scaling limit. In this field theory
approach however strings are not allowed to have different tensions,
which is a characteristic property of cosmic superstrings.  This issue
has been addressed later in the context of modelling $(p,q)$ cosmic
superstrings~\cite{csn4}. It was found that such networks rapidly
approach a stable scaling solution, where once scaling is reached,
only a small number of the lowest tension states is populated
substantially. An interesting question is to find out whether the
field theory approach of Ref.~\cite{csn3} mimics the results of the
modelling approach of Ref.~\cite{csn4}. Finally, performing full
classical field theory simulations for a model of a string network
with junctions, where the junctions can be thought of as global
monopoles connected by global strings, it was shown~\cite{csn6} that
the evolution is consistent with a late-time scaling regime. Thus,
the presence of junctions is not itself inconsistent
with scaling.

The cosmic superstring network is characterised~\cite{csn1} by two
components: there are a few long strings with a scale-invariant
evolution; the characteristic curvature radius of long strings, as
well as the typical separation between two long strings are both
comparable to the horizon size, $\xi(t)\simeq {\sqrt {\cal P}} t$, and
there is a large number of small closed loops having sizes $\ll t$.
Assuming there are string interactions, the network of long strings
will reach an asymptotic energy density, where the energy density in
long strings is
\begin{equation}
\rho_{\rm l}=\frac{\mu}{{\cal P} t^2}~.
\end{equation}
Thus, the fraction of the total density in the form of strings in the
radiation-dominated era reads
\begin{equation}
\frac{\rho_{\rm str}}{\rho_{\rm total}}=\frac{32\pi}{ 3}
\frac{G\mu}{{\cal P}}~.
\end{equation}

Recent numerical investigations~\cite{csn5} of strings evolving in a
matter- or radiation-dominated FLRW background claim a weaker power
law for the dependence of the scaling string energy density. More
precisely, in Ref.~\cite{csn5} it was found that for ${\cal P}\gsim
0.1$, the function $\rho(1/{\cal P})$ is approximately flat, while for
${\cal P}\lsim 0.1$, the function $\rho(1/{\cal P})$ is well-fitted by
a power-law with exponent $0.6^{+0.15}_{-0.12}$.  The behaviour of the
string energy density as a function of ${\cal P}$ has an important
impact for the observational consequences of cosmic superstring
networks. 

Oscillating string loops loose energy by emitting graviton, dilaton
and Ramond-Ramond (RR) fields. Accelerated cosmic strings are sources
of gravitational radiation, in particular from the vicinity of the
cusps where the string velocity approaches the speed of
light. Similarly, cosmic superstrings emit gravity waves but since the
intercommutation probability is less than unity, their network is
denser with more cusps, resulting in an enhancement of the emitted
gravitational radiation. As it was pointed out~\cite{dv}, the
gravitational wave bursts emitted from cusps of oscillating string or
superstring loops could be detectable with the gravitational-wave
interferometers LIGO/VIRGO and LISA.  

One can place constraints on the energy scale of cosmic strings from
the observational bounds on dilaton decays~\cite{tdv}.  Considering
that the dilaton lifetime is in the range $10^7{\rm s}\lsim \tau\lsim
t_{\rm dec}$, one can obtain an upper bound $\eta\lsim {\cal
P}^{-1/3}\lsim 10^{11}{\rm GeV}$~\cite{csn1} for the energy scale of
cosmic superstrings, which determines the critical temperature for the
transition leading to string formation. A lower reconnection
probability allows a higher energy scale of strings, at most by one
order of magnitude.

\section{Conclusions}

A realistic cosmological scenario necessitates the input of high
energy physics, implying that models describing the early stages of
the evolution of the Universe have their foundations in both general
relativity as well as high energy physics. Comparing the predictions
of such models against current astrophysical and cosmological data one
concludes to either their acceptance or their rejection, while in the
first case one can also fix the free parameters of the models.  One of
the most beautiful examples in this interplay between cosmology and
high energy phyics is the case of cosmic strings.

Cosmic strings are expected to be generically formed during the
evolution of the Universe, provided the general theoretical picture we
have in mind is correct. However, many independent studies concluded
in the robust statement that cosmic strings have a limited r\^ole in
the measured CMB temperature anisotropies. Knowing the upper bounds on
the contribution of strings to the CMB, one has to examine whether the
theoretical models can be adjusted so that there is an agreement
between predictions and data. This issue has been addressed in length
here. In this respect, cosmology uses high energy physics to build a
natural and successful cosmological model, while it offers back some
means for testing high energy physics itself.

Cosmic strings are a robust prediction of GUTs, or even M-theory.
Even though their r\^ole in explaining the origin of the observed
large-scale structure is sub-dominant, their astrophysical and
cosmological implications remain important. Cosmic strings are a small
but by no means negligible contribution to any successful cosmological
model.

%

\end{document}